\documentclass[12pt,preprint]{aastex}

\slugcomment{}

\shorttitle{Compact Flare}
\shortauthors{Joshi et al.}

\begin{document}

%------------------------------------------------------------------------------------------

\title{Generalisation of the Magnetic Field Configuration of typical and atypical Confined Flares}

%------------------------------------------------------------------------------------------

\author{Navin Chandra Joshi\altaffilmark{1,2}, Xiaoshuai Zhu\altaffilmark{3}, Brigitte Schmieder\altaffilmark{4,2}, Guillaume Aulanier\altaffilmark{4}, Miho Janvier\altaffilmark{5}, Bhuwan Joshi\altaffilmark{1}, Tetsuya Magara\altaffilmark{2}, Ramesh Chandra\altaffilmark{6}, Satoshi Inoue\altaffilmark{7}}

\altaffiltext{1}{Udaipur Solar Observatory, Physical Research Laboratory, Udaipur 313 001, India}
\altaffiltext{2}{School of Space Research, Kyung Hee University, Yongin, Gyeonggi-Do, 446-701, Republic of Korea}
\altaffiltext{3}{Max-Planck Institute for Solar System Research, Goettingen 37077, Germany}
\altaffiltext{4}{LESIA, Observatoire de Paris, PSL Research University, CNRS Sorbonne Universit\'e, Univ. Paris 06, Univ. Paris Diderot, Sorbonne Paris Cit\'e, 5 place Jules Janssen, F-92195 Meudon, France}
\altaffiltext{5}{Institut d'Astrophysique Spatiale, CNRS, Univ. Paris-Sud, Universit\'e Paris-Saclay, Bât. 121, 91405 Orsay CEDEX, France}
\altaffiltext{6}{Department of Physics, Kumaun University, DSB Campus, Nainital 263001, India}
\altaffiltext{7}{Institute for Space-Earth Environmental Research, Nagoya University, Furo-cho, Chikusa-ku, Nagoya, 464-8601, Japan}

%------------------------------------------------------------------------------------------

\begin{abstract}

Atypical flares cannot be naturally explained with standard models. To predict such flares, we need to define their physical characteristics, in particular, their magnetic environment, and identify pairs of reconnected loops. Here, we present in detail a case--study of a confined flare preceded by flux cancellation that leads to the formation of a filament. The slow rise of the non--eruptive filament favours the growth and reconnection of overlying loops. The flare is only of C5.0 class but it is a long duration event. The reason is that it is comprised of three successive stages of reconnection. A non--linear force--free field extrapolation and a magnetic topology analysis allow us to identify the loops involved in the reconnection process and build a reliable scenario for this atypical confined flare.  The main result is that a curved magnetic polarity inversion line in active regions is a key ingredient for producing such atypical flares. A comparison with previous extrapolations for typical and atypical confined flares leads us to propose a cartoon for generalizing the concept.

\end{abstract}

%------------------------------------------------------------------------------------------

\keywords{Sun: Flare - Sun: Magnetic Reconnection - Sun: Magnetic Field}

%------------------------------------------------------------------------------------------

\section{Introduction}
\label{sec1}

Solar flares are the manifestation of the violent energy release taking place in the solar corona via magnetic reconnection. The energy powering these flares is stored in the highly sheared, current--carrying magnetic field lines in the corona. Then, instabilities and photospheric motions can lead to the sudden release of this energy via the reconnection of coronal magnetic field lines \citep{Shibata11}. The initial configuration of the coronal field significantly influences how the flare proceeds. Various categories of flares exist: eruptive flares, flares with partial or failed eruptions, and confined (or compact) flares. Flares with full or partial eruptions produce coronal mass ejections (CMEs) while flares with failed eruption and compact flares are not associated with CMEs \citep[][and references therein]{Schmieder13}.

Over the years, many observational studies have investigated the physical conditions that determine the evolution of eruptive and confined flares. These studies can be classified into two different categories, based on statistical or case--study approach. Both approaches have their own merit. With statistics, reliable parameters have been determined to explain the occurrence of  energetic flares. By exploring the connection between the photospheric magnetic field and solar flares,  it was found that there exists a strong correlation between non--potentiality and flare strength.  Different parameters have been  used in such studies e.g., the length of the polarity inversion line (PIL), the strong shear of the transverse magnetic field, the degree of non--potentiality of the active region, as well as the decay index \citep{Hagyard1990,Leka1993,Falconer2001,Falconer2003,Abramenko2005,Schrijver2007,Joshi14a,Zuccarello2015,Jos14b}. For example, the recent study of \citet{Vasantharaju18} shows statistically a good correlation between non--potentiality and flare strength. In this study, a total of 77 cases were studied, with the equal number of events in both confined and eruptive categories. However, the study did not show a clear parameter to predict which flare would be eruptive or compact. While a statistical analysis provides probabilities that may be indicative of flare characteristics, such a study does not provide clarity about the underlying physical processes, e.g., the triggering mechanism and the magnetic configuration of reconnection.

The case--studies allow us to investigate the physical mechanisms responsible for reconnection and eruption. The case--study approach is complementary to statistical studies, which aim at identifying macroscopic and global tendencies. The statistical approach has helped in building several models of solar flares. For example, the two--dimensional (2D) model of flare known as the ``CSHKP" model \citep{Carmichael64,Sturrock66,Hirayama74,Koop76} or the three--dimensional (3D) standard flare model \citep[see,][]{Aulanier12,Janvier15}, explain that flares are the consequences of reconnection between magnetic field lines, in particular, the legs of surrounding arcades, below the eruptive filament or/and flux rope  \citep{Aulanier12}. But it is the case--study approach that helped to build other scenarios such as the 3D ``tether cutting model" \citep{Moore01}, or testing the so--called breakout model of \citet{Antiochos99} \citep{Aulanier2000,Ugarte2007}. Departure from such so--called ``standard" models provides the possibility to better understand other scenarios. 

\begin{figure}[!ht]
\vspace*{-4cm}
\centerline{
	\hspace*{0.0\textwidth}
	\includegraphics[width=3\textwidth,clip=]{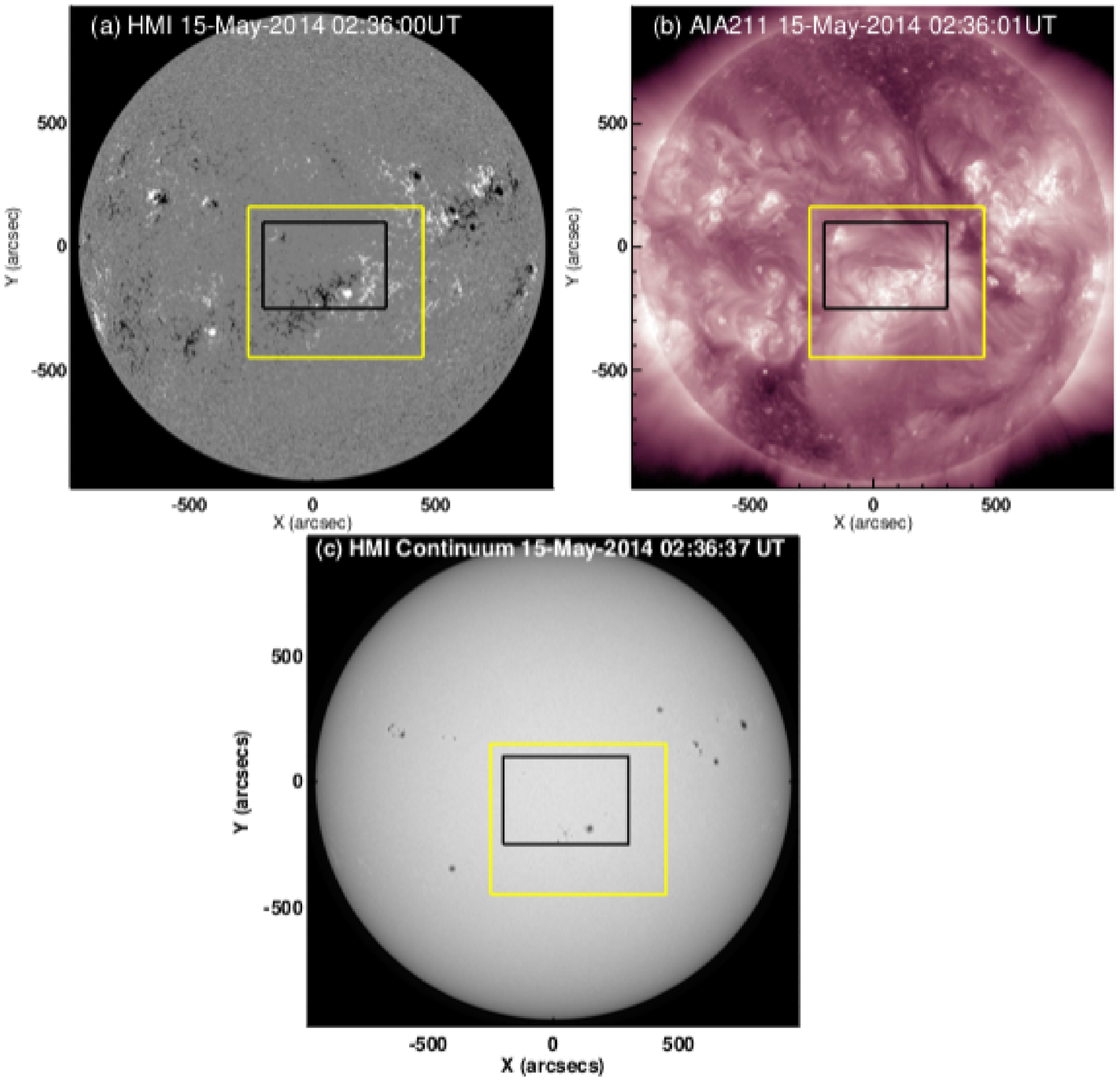}
	}
\vspace*{-7cm}
\caption{(a) \textit{SDO}/HMI photospheric magnetogram on 2014 May 15 at 02:36:00 UT. (b) \textit{SDO}/AIA 211 \AA\ image on 2014 May 15 at 02:36:01 UT. (c) \textit{SDO}/HMI continuum image at 02:36:37 UT on 2014 May 15. The yellow box shows the field of view of Figure~\ref{fig13} including ARs 12058 and 12060 and the black box shows the field of view of Figure~\ref{fig3}.}
\label{fig1}
\end{figure}

The case study approach has revealed, in particular, the importance of magnetic field environments. Magnetic field configuration of active regions is often the key to explain eruptions and flares \citep[e.g.,][]{Mandrini2014,Masson09,Joshi15}. These studies point towards the importance of a null point above an emerging flux region in the pre--existing fan--spine magnetic field configuration. However, null points are not always efficient for driving eruptive flares as shown by the work of \citet{Zuccarello2017} which presents such a case with a null point. They investigated 10 flares from a single AR and found that during
the first day flares were eruptive, while the events occurring on the next day were confined. The only possibility that they found to explain this different behaviour was the respective orientation of the emerging field lines with that of the overlying field.

More generally the analysis of the topology of an active region leads to information about the regions where reconnection is possible. These regions are  3D fine layers where reconnection can occur because the connectivity of magnetic field lines can change drastically. These regions are called quasi--separatrix layers (QSLs) \citep{Demoulin96,Demoulin97}. During flares, strong electric currents develop in  QSLs. Flare ribbons correspond to the QSL footprints in the photosphere. With the QSL analysis, one may able to understand which magnetic field lines may reconnect.

The identification of QSLs is possible via magnetic field extrapolation using observed magnetograms of the photosphere. Confined flares are usually studied by this method. The QSL footprints in the photosphere correspond to multiple ribbons in compact flares \citep{Mandrini96,Chandra09,Dalmasse15}. In fact, there exists a series of atypical flares that do not match with either eruptive, null point or usual anti--parallel QSL models  \citep{Schmieder97,Dalmasse15}. Only a topological analysis allows us to understand the  3D magnetic configurations and propose convincing scenarios of reconnection leading to such atypical flare.

\begin{figure}[!ht]
\vspace*{-2cm}
\centerline{
	\hspace*{0.0\textwidth}
	\includegraphics[width=1.5\textwidth,clip=]{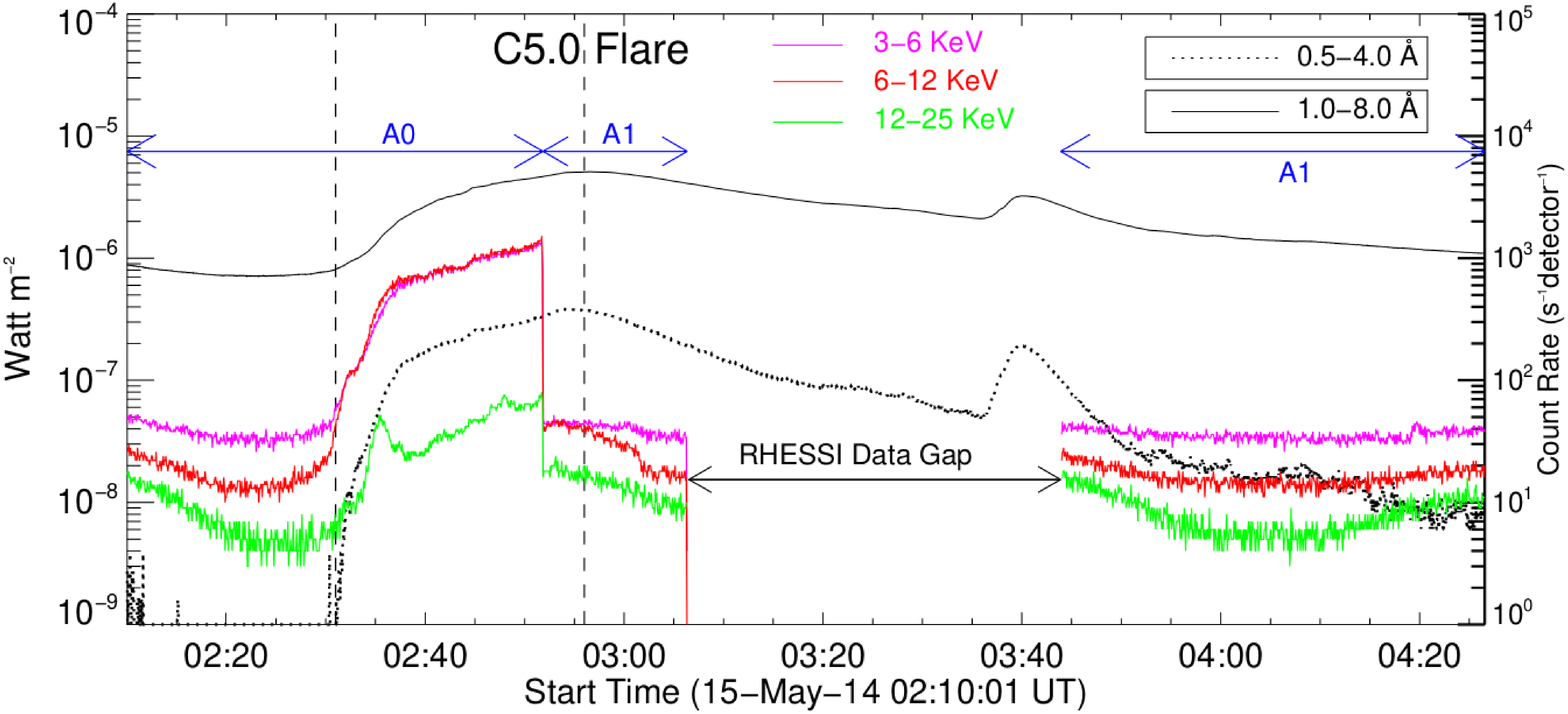}
	}
\vspace*{-3cm}
\caption{GOES and RHESSI X--ray profiles between 02:10:01 UT and 04:26:33 UT on 2014 May 15. The vertical lines from left to right represent the start ($\approx$02:31 UT) and the peak ($\approx$02:56 UT) times of the flare, respectively. Horizontal bars represent different RHESSI's attenuator states (A0 and A1).}
\label{fig2}
\end{figure}

In this article, we study an atypical confined flare of GOES class C5.0 that leads to a long duration event (LDE) in soft X--rays (SXR) and provides a topological model to understand such exceptional cases. By carefully studying multi--wavelength data set, we aim to decipher the physical mechanism for this event and draw comparisons with previous cases \citep{Schmieder97,Dalmasse15}. By doing so, our objective is to generalize the conditions that produce such atypical flares. The structure of the paper is as follows.  Section~\ref{sec2} describes the instruments from which the data set come.  Section~\ref{sec3}  describes the characteristics of flaring activity region, its magnetic configuration and the dynamic evolution of the flare. In Section~\ref{sec4}, the magnetic field modeling is presented that include a non--linear force--free field (NLFFF) magnetic field extrapolation (Sub--section~\ref{sec4.1}), QSL computation (Sub--section~\ref{sec4.2}) and their comparison with multi--wavelength imaging observations (Sub--section~\ref{sec4.3}). Results and generalization of magnetic configuration for compact/confined flares are discussed in Section~\ref{sec5}.

%------------------------------------------------------------------------------------------

\section{Observational data set}
\label{sec2}

We used a multi--wavelength data set to analyze this event. The analysis is primarily based on the observations taken by the \textit{Atmospheric Imaging Assembly} \citep[AIA;][]{Lem12}, on board the \textit{Solar Dynamics Observatory} \citep[\textit{SDO};][]{Pesnell12}. AIA observes the Sun in seven extreme ultraviolet (EUV) wavelength channels, two ultraviolet (UV) and one white light channel. The AIA takes images in EUV with a cadence of 12 s, UV with a cadence of 24 s and white light channel with a cadence of 1 hr having a pixel size of  $\rm 0.6\arcsec$. The images from the hottest AIA channels
($\rm 131~\AA, T\sim 10^{7} K; 94~\AA, T\sim 10^{6} K$) 
are used to analyze the flare and the evolution of flare loops. We further explore data from AIA 304 \AA\ (T$\sim$50,000 K) and 1600 AA\ ($\rm T\sim 10^{5} K$) data to analyze the flare ribbon dynamics. It is noteworthy that the AIA 1600~\AA\ bandpass includes a part of the continuum formed in the temperature minimum region at the temperature of T$\sim$5000 K, as well as the C IV doublet at 1550 \AA\ formed in the transition region of ($\rm T\sim 10^{5} K$) \citep{Brekke1994}. At flare ribbons, emission in this passband is significantly enhanced largely due to increased C IV line emissions. For the magnetic field analysis, we used data from the \textit{Helioseismic and Magnetic Imager} \citep[\textit{HMI};][]{Schou12} onboard the \textit{SDO}. The observational cadence of this instrument is 45 s (for the line--of--sight photospheric magnetogram) and 720 s (for the vector magnetic field data) and the pixel size of the images is $\rm 0.5\arcsec$. The flare - associated X--ray sources have been identified with the observations made by the \textit{Reuven Ramaty High Energy Solar Spectroscopic Imager} \citep[\textit{RHESSI};][]{Lin12}. For the filament detection, we used the observations provided by the spectroheliograph of the Observatoire de Paris in Meudon which are archived in the database of the solar survey\footnote{$\rm http://bass2000.obspm.fr/home.php?lang=fr$}.
Full disk spectroheliograms in $\rm H\alpha$ and Ca H lines are obtained for several wavelengths in each line in a daily cadence.

\begin{figure}[!ht]
\vspace*{-8cm}
\centerline{
	\hspace*{0.0\textwidth}
	\includegraphics[width=2.5\textwidth,clip=]{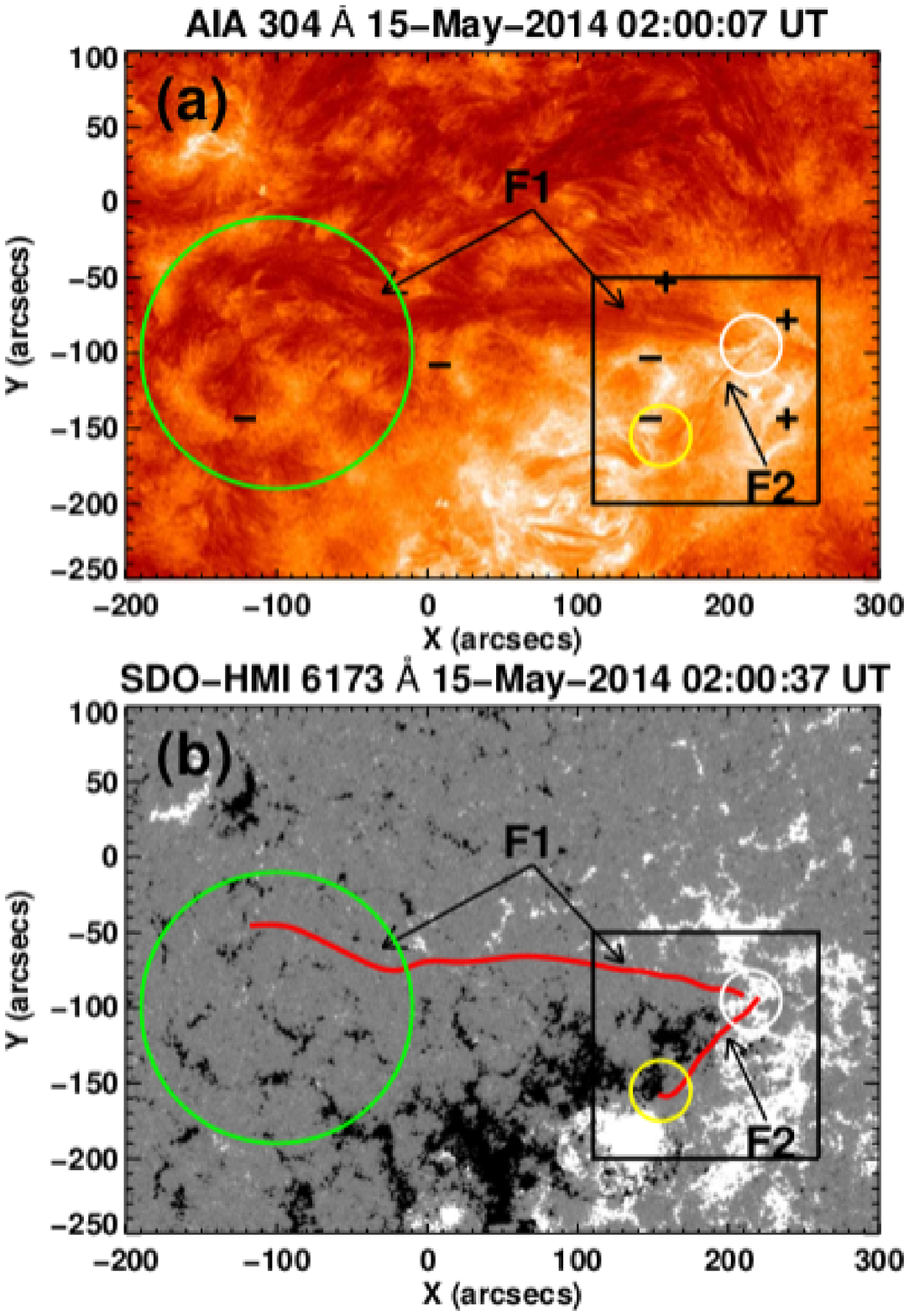}
	}
\vspace*{-5cm}
\caption{Upper panel (a): \textit{SDO}/AIA 304 \AA\ image on 2014 May 15 at 02:00:07 UT, showing the northern (F1) and southern (F2) filaments. The white circle represents the area where the north--west footpoint of filament F2 and the western footpoint of filament F1 is anchored. The green and yellow circles show the eastern footpoint of F1 and south--east footpoint of F2, respectively. Bottom panel (b): \textit{SDO}/HMI line--of--sight photospheric magnetogram on 2014 May 15 at 02:00:37 UT. The overplotted red lines represent the center axes of the filaments. These red lines are traced from AIA 304 \AA\ image shown in panel (a). The black box indicates the field of view of Figure~\ref{fig5}(c).}
\label{fig3}
\end{figure}

%------------------------------------------------------------------------------------------

\section{Multi--wavelength observations and analysis}
\label{sec3}

\subsection{General configuration of the active region} 
\label{sec3.1}

The flaring activity concerns two NOAA active regions (ARs) namely 12058 and 12060. AR 12058 consists of elongated plages or intense network areas. The AR 12060 appeared in the middle of the AR 12058. The magnetic classes of the ARs 12058 and 12060 were $\beta$ and $\beta \gamma$, respectively on the day of the event. The ARs lie near the solar center with average positions S10W26 (AR 12058) and S14W19 (AR 12060).
Both ARs have a leading region with a strong positive polarity and the following region with dispersed negative polarities (Figure~\ref{fig1}).

\begin{figure}[!ht]
\vspace*{-6cm}
\centerline{
	\hspace*{0.0\textwidth}
	\includegraphics[width=2.2\textwidth,clip=]{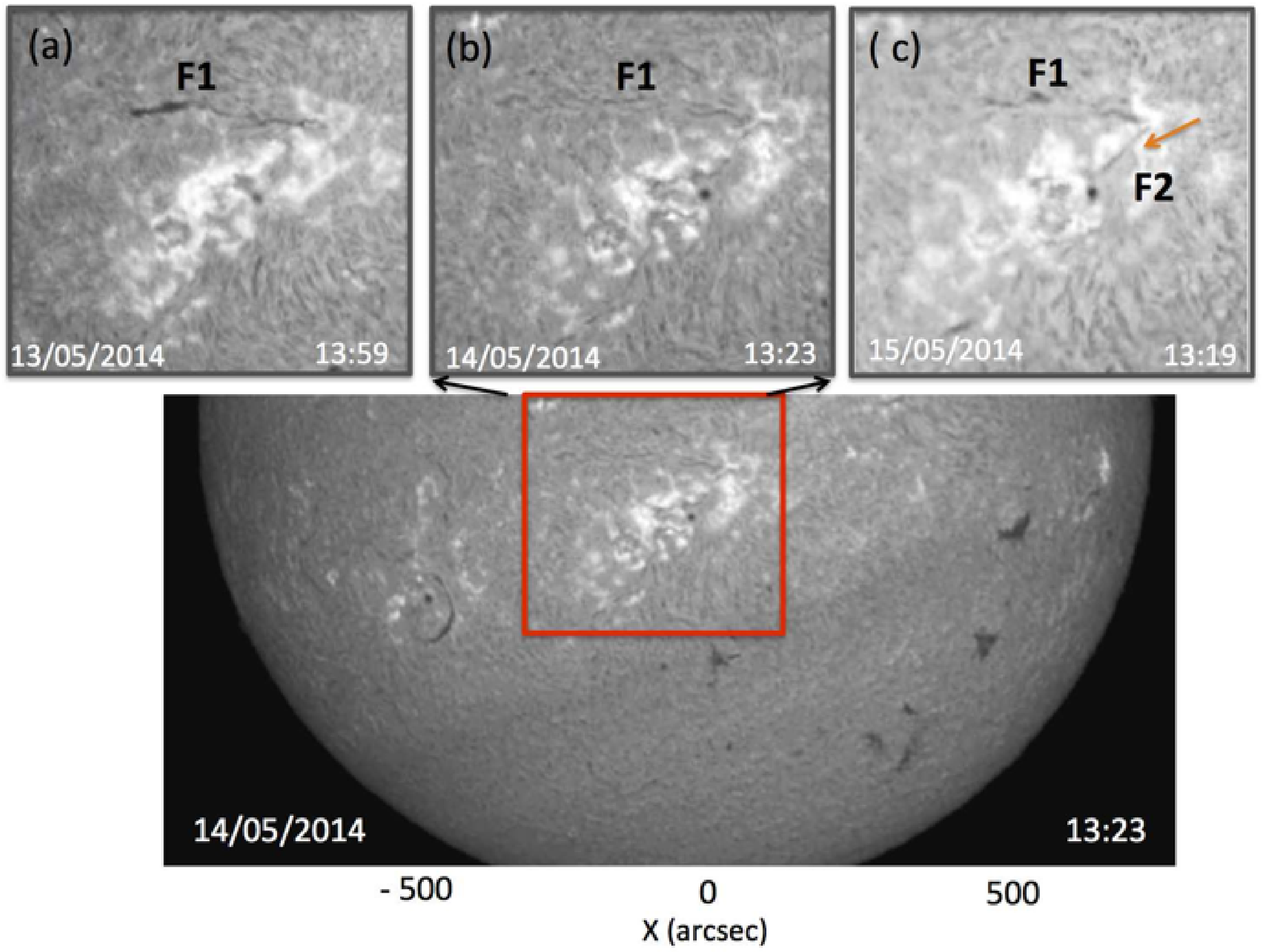}
	}
\vspace*{-5cm}
\caption{$\rm H\alpha$ spectroheliograms of the Meudon survey on 2014 May 13, 14, and 15, showing the formation of the filament F2.}
\label{fig4}
\end{figure}

A long duration C5.0 class flare occurred on 2014 May 15 which persisted for $\approx$2 hr. The temporal evolution of GOES and RHESSI fluxes are shown in Figure~\ref{fig2}. The GOES SXR profiles (shown in black colour) show that the flare started $\approx$ 02:31 UT, reached its peak in a gradual manner at $\approx$ 02:56 UT. It underwent a long decay phase until $\approx$ 04:30~UT. The RHESSI X--ray profiles are shown by different coloured lines in Figure~\ref{fig2}. The flare occurred in the northern part of the two active regions (S08W26), where two filaments F1 and F2 are observed in the AIA  304~\AA\ image  (Figure~\ref{fig3}(a)). F1 is well visible as a broad dark elongated area in 211 \AA\ (Figure~\ref{fig1}(b)) and in 304 \AA\ (Figure~\ref{fig3}(a)) images, while  F2  is visible as a narrow dark lane  only in 304 \AA\  image (Figure~\ref{fig3}(a)). We traced the center line of both filaments observed in 304 \AA\ and overplotted on the HMI photospheric magnetogram (red lines in   Figure~\ref{fig3}(b)). The filaments correspond to the two main polarity inversion lines (PIL) that make an angle of $\approx$45 degrees from each other. We will show that the curvature of the composite PIL is a key to understand the flare build--up.

\begin{figure}[!ht]
\vspace*{-6cm}
\centerline{
	\hspace*{0.0\textwidth}
	\includegraphics[width=2.8\textwidth,clip=]{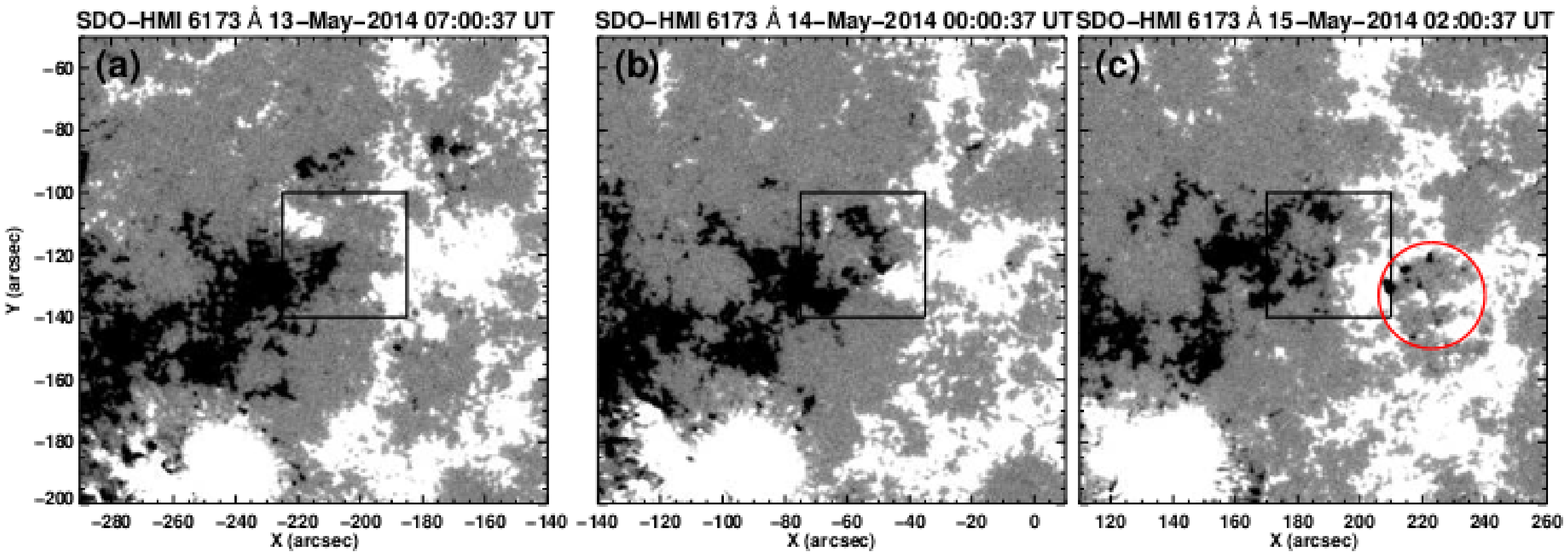}
	}
\vspace*{-10cm}
\caption{\textit{SDO}/HMI line--of--sight photospheric magnetograms on 2014 May 13, 14, and 15. The area shown by the boxes in each panel is the flux cancellation region during the formation of the filament F2. The region shown by the red circle in panel (c) is the region of flux emergence a few hours before the flare onset. The field of view is $\rm 150\arcsec$ by $\rm 150\arcsec$.}
\label{fig5}
\end{figure}

The large--scale evolution of the filaments during three days before the flare is presented in panels (a)--(c) of Figures~\ref{fig4} which are the zoomed images taken from the daily H$\alpha$ spectroheliograms (Figure~\ref{fig4}). F1 was already present on May 13 while F2 was gradually forming between May 14 and May 15. It can be clearly seen (Figure~\ref{fig3}(b)) that the western ends of both the filaments, F1 and F2, are anchored at the same positive polarity region (indicated by the white circle), implying  the junction of the two PILs. The other end of F1 lies in a negative region of weak and dispersed magnetic field (large green circle in Figure~\ref{fig3}(b)) while the other end of F2 is at the negative region in the north of the major positive polarity (yellow circle in Figure~\ref{fig3}(b)).

%------------------------------------------------------------------------------------------

\subsection{Magnetic field changes and formation of filament F2}
\label{sec3.2}

The photospheric magnetic field changes can be seen in Figure~\ref{fig5} that present the zoomed images of the region where the two filaments have their common end in the positive polarity  (white circle of Figure~\ref{fig3}(b)). On the left of the positive polarity, we see the evolution of negative polarity structures that seem to approach towards the positive polarity region over the period of three days before the flare, which cancels the existing positive polarity. We further note the flux emergence within the positive polarity region. Both changes are shown in Figure~\ref{fig5}(c) by a black box and a red circle, respectively. They can be followed  in the SDO/HMI movie attached with Figure~\ref{fig5}. Small and continuous cancellations of magnetic polarities and emerging flux can push the existing positive polarities towards the PIL. Such converging motions could be the cause of the gradual growth of the filament F2 as observed between May 13 and May 15 (Figure~\ref{fig4}(a)--(c)). This growth of F2 suggests that the magnetic loops that are surrounding F2 grow and naturally rise as proposed in flux cancellation models of filament formation \citep{vanBallegooijen89,Aulanier10}. 

\begin{figure}[!ht]
\vspace*{-9cm}
\centerline{
	\hspace*{0.03\textwidth}
	\includegraphics[width=3\textwidth,clip=]{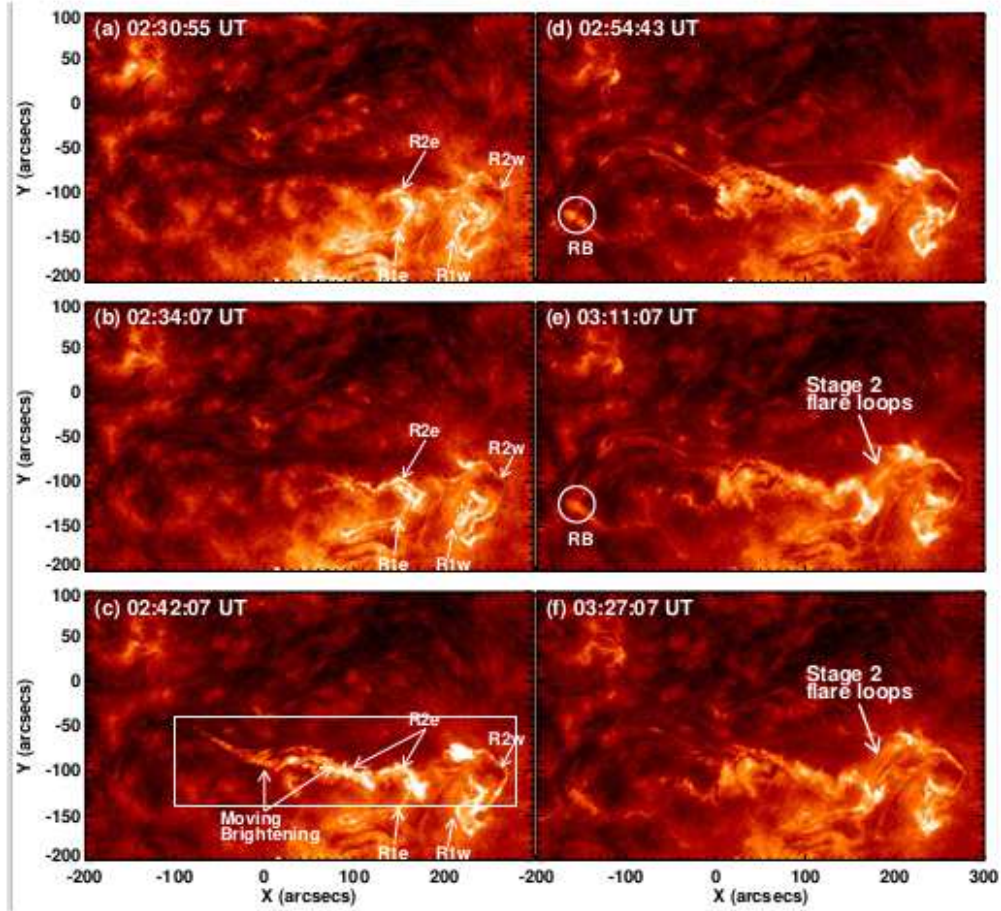}
	}
\vspace*{-7.5cm}
\caption{The sequence of selected \textit{SDO}/AIA 304 \AA\ images showing the overall dynamics of the formation of different sets of flare ribbons and associated brightenings. White circle in panels (d) and (e) represent the area of the remote brightening (RB). The white box in panel (c) represents the field of view of Figure~\ref{fig7}(a).}
\label{fig6}
\end{figure}

%------------------------------------------------------------------------------------------

\subsection{Dynamics evolution of the confined solar flare}
\label{sec3.3}

%--------------------------------------------

\begin{figure}[!ht]
\vspace*{-2cm}
\centerline{
	\hspace*{-0.02\textwidth}
	\includegraphics[width=1.9\textwidth,clip=,]{fig7.eps}
	}
\vspace*{-2.5cm}
\caption{(a) A cutout of the \textit{SDO}/AIA 304 \AA\ images, showing the moving brightness along the ribbon R2e. The dashed line show the cut along which the distance--time map has been made. (b) The distance--time map of the moving brightness. The speed of the moving brightness is $\rm \approx 200~km~s^{-1}$. The field of view of the panel (a) is shown in Figure~\ref{fig6}(c) by the white rectangle.}
\label{fig7}
\end{figure}

During the long--duration of the flare, three different phases can be identified according to the evolution of the sets of flare ribbons and loops system that provide evidence for three phases of reconnection. We call these three phases as stage 1,  stage 2, and stage 3, respectively.

%--------------------------------------------

\subsubsection{Flare ribbons}
\label{sec3.3.1}

The dynamic evolution of the flare ribbons can be seen in AIA 304 \AA\ images (Figure~\ref{fig6}) and associated movie. Two pairs of flare ribbons on each side of the filament F2 (R1e/R1w and R2e/R2w) are clearly visible at $\approx$02:30 UT. The letters ``e" and ``w" indicate the eastern and the western sides of F2, respectively. At $\approx$02:30 UT, R2e and R2w were faint, while by $\approx$02:42 UT  they become bright and rapidly extend. These two sets of ribbons correspond to the first two stages of the flare.

\begin{figure}[!ht]
\vspace*{-9cm}
\centerline{
	\hspace*{0.03\textwidth}
	\includegraphics[width=3\textwidth,clip=]{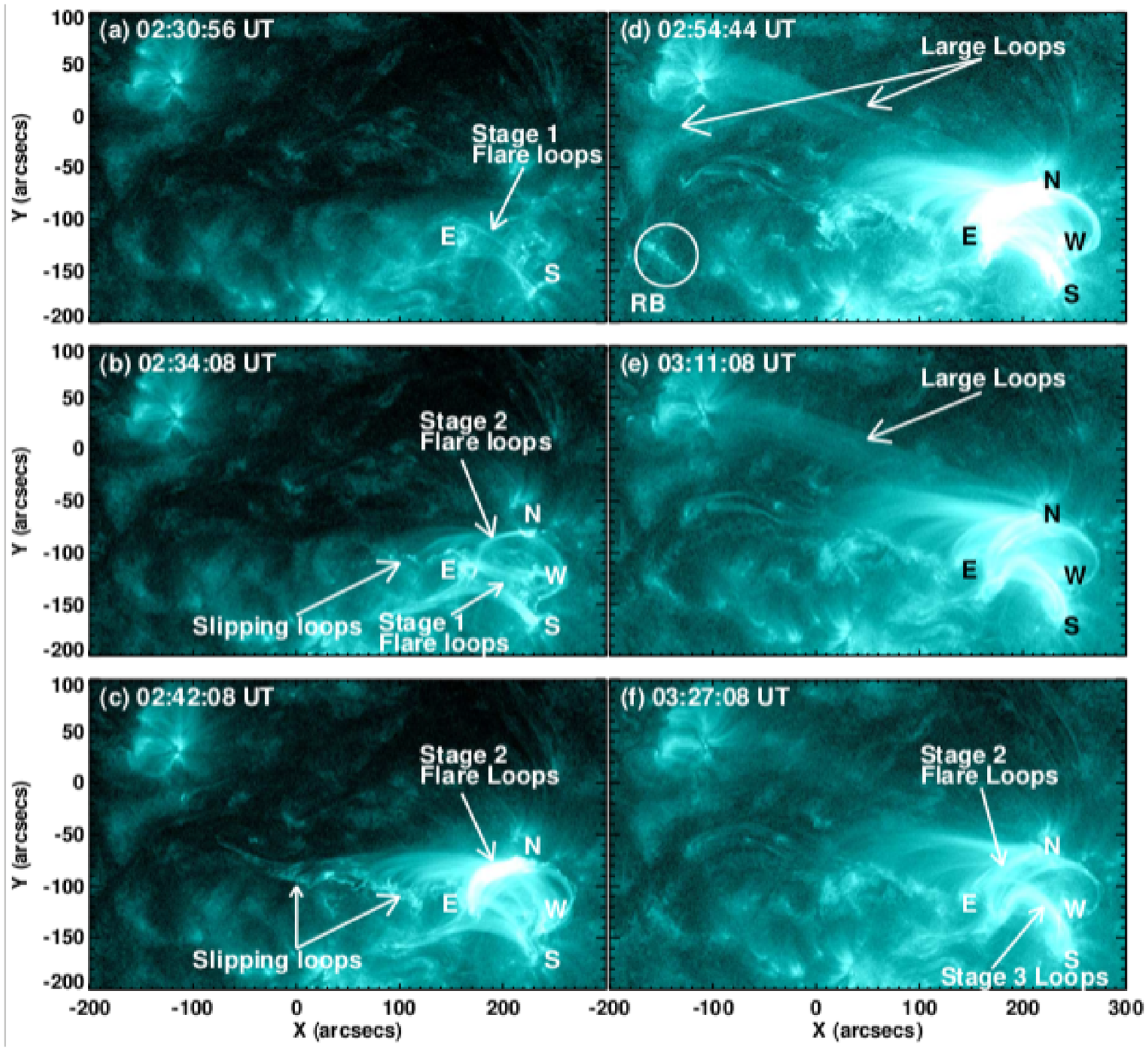}
	}
\vspace*{-6.5cm}
\caption{The sequence of selected \textit{SDO}/AIA 131 \AA\ images, showing the overall dynamics of the flare loops in the coronal region. In panel (a), we see the flare loops forming by the reconnection during the stage 1 of the event. In panel (b), along with the flare loops of stage 1, we also see the formation of new set of flare loops forming during the stage 2 of the event, which became brighter with time (as seen in panel (c)). The remote brightening (RB) area is marked by the white circle in panel (d). The post--flare loops of stage 2 and the newly formed flare loops of stage 3 of the event can be seen in panel (f).}
\label{fig8}
\end{figure}

We observe bright kernels moving along the ribbon R2e (Figure~\ref{fig6}(c)). Such slipping kernels  have been studied in detail as the subsequent motion of flare loops following 3D reconnection in \cite{Dudik14} (c.f. Figures~\ref{fig6}(b)--(c) and~\ref{fig8}(b)--(c)).  
The speed measurement of the kernels along the flare ribbon in 304 \AA\ is shown in Figure~\ref{fig7}. Figure~\ref{fig7}(b) shows the time--distance plot of the moving brightness along the trajectory shown by the white dashed line in Figure~\ref{fig7}(a). The average speed of kernels along the ribbon R2e is $\approx$200~km~s$^{-1}$. According to \citet{Dudik14}  the propagation of brightening along a flare ribbon can be attributed to the rapid change of connectivity that flare loops undergo during the reconnection process. This mechanism is also referred to as slipping or slip--running loops due to the intrinsic nature of reconnection in 3D \citep[see][]{Aulanier06}. These complicated multiple ribbons do not allow us to understand the magnetic reconnection during the flare with clarity. For that, we had to investigate the evolution of the bright loops seen in 131 \AA\ and 94 \AA\ images of AIA.

%--------------------------------------------

\subsubsection{Loops in stage 1}
\label{sec3.3.2}

\begin{figure}[!ht]
\vspace*{-7cm}
\centerline{
	\hspace*{0.0\textwidth}
	\includegraphics[width=2.8\textwidth,clip=]{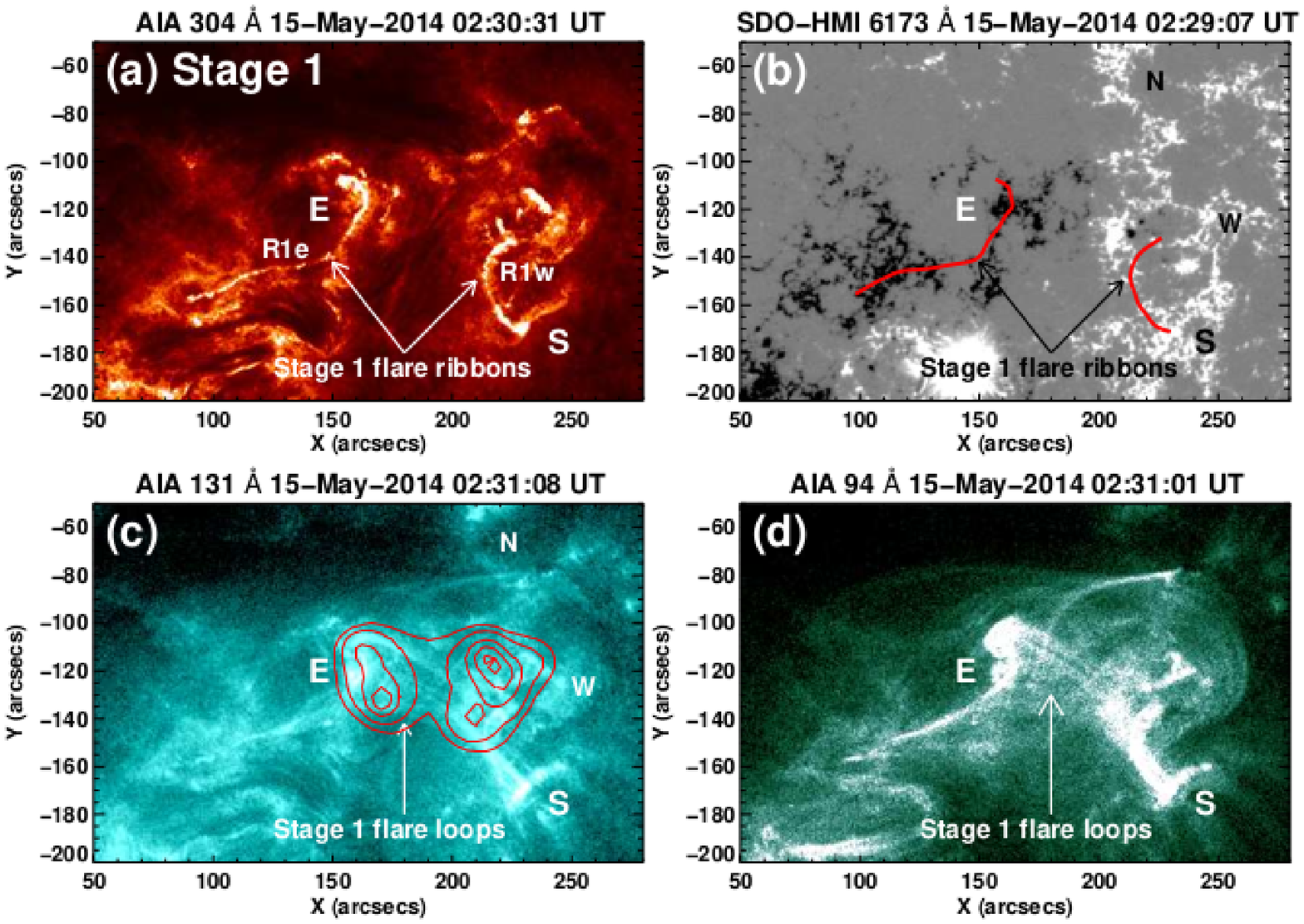}
	}
\vspace*{-7cm}
\caption{(a) \textit{SDO}/AIA 304 \AA\ image showing the flare ribbons formed during the stage 1 of the flare. (b) \textit{SDO}/HMI line--of--sight photospheric magnetogram at 02:29:07 UT on 2014 May 15. The overplotted red lines show the flare ribbons locations. These red lines are traced from the AIA 304 \AA\ image shown in panel (a). \textit{SDO}/AIA 131 \AA\ and 94 \AA\ images showing the flare loops connecting the flare ribbons are shown in the panels (c) and (d), respectively. The red contours in panel (c) are the RHESSI X--ray contours of 6--12 keV energy band. The contours levels are the 15\%, 25\%, 50\%, 75\%, and 95\% of the peak intensity.}
\label{fig9}
\end{figure}

Figure~\ref{fig8} represents the sequence of selected images in the AIA 131 \AA\ channel, showing the development of flare loop systems. In stage 1, we find bright loops joining footpoint regions E (east) and S (south) in the south--west of the region at $\approx$02:30 UT (Figure~\ref{fig8}(a)). At $\approx$02:34 UT (Figure~\ref{fig8}(b)) a system of bright reconnected loops joining E and S are clearly visible. The multi--wavelength view of stage 1 is depicted in Figure~\ref{fig9}. The  R1e and R1w ribbons are over--plotted on the \textit{SDO}/HMI photospheric magnetogram to investigate their locations  compared with the overall magnetic field (Figure~\ref{fig9}(b)). It can be seen that the eastern (R1e) and western (R1w) ribbons lie on the negative and positive polarities, respectively. The set of loops observed in the hottest channels of AIA (i.e., 94 and 131 \AA) are the post--reconnected loop system that joins the flare ribbons (Figure~\ref{fig9}(c)--(d)).

%--------------------------------------------

\subsubsection{Loops in stage 2}
\label{sec3.3.3}

At $\approx$02:34 UT a  new set of reconnected bright loops joining regions E and N (E and N mark the  eastern and the northern sites of the flaring regions, respectively) starts to appear (Figure~\ref{fig8} (b)). These loops define the beginning of reconnection in stage 2. By $\approx$03:27 UT, bright post--flare loops develop over the region where both the filaments, F1 and F2, are embedded (Figure~\ref{fig8}(f)). It is noteworthy that large loops are observed connecting the western and eastern part of the active region (marked by arrows in Figure~\ref{fig8}(d) and~\ref{fig8}(e)), with a remote brightening (RB) clearly visible at $\approx$02:54 UT, indicating that these long loops have some relationships with the flare. Notably, the RB structure is also confirmed by the 304 \AA\ (Figure~\ref{fig6}(d) and~\ref{fig6}(e)).

\begin{figure}[!ht]
\vspace*{-7cm}
\centerline{
	\hspace*{0.0\textwidth}
	\includegraphics[width=2.5\textwidth,clip=]{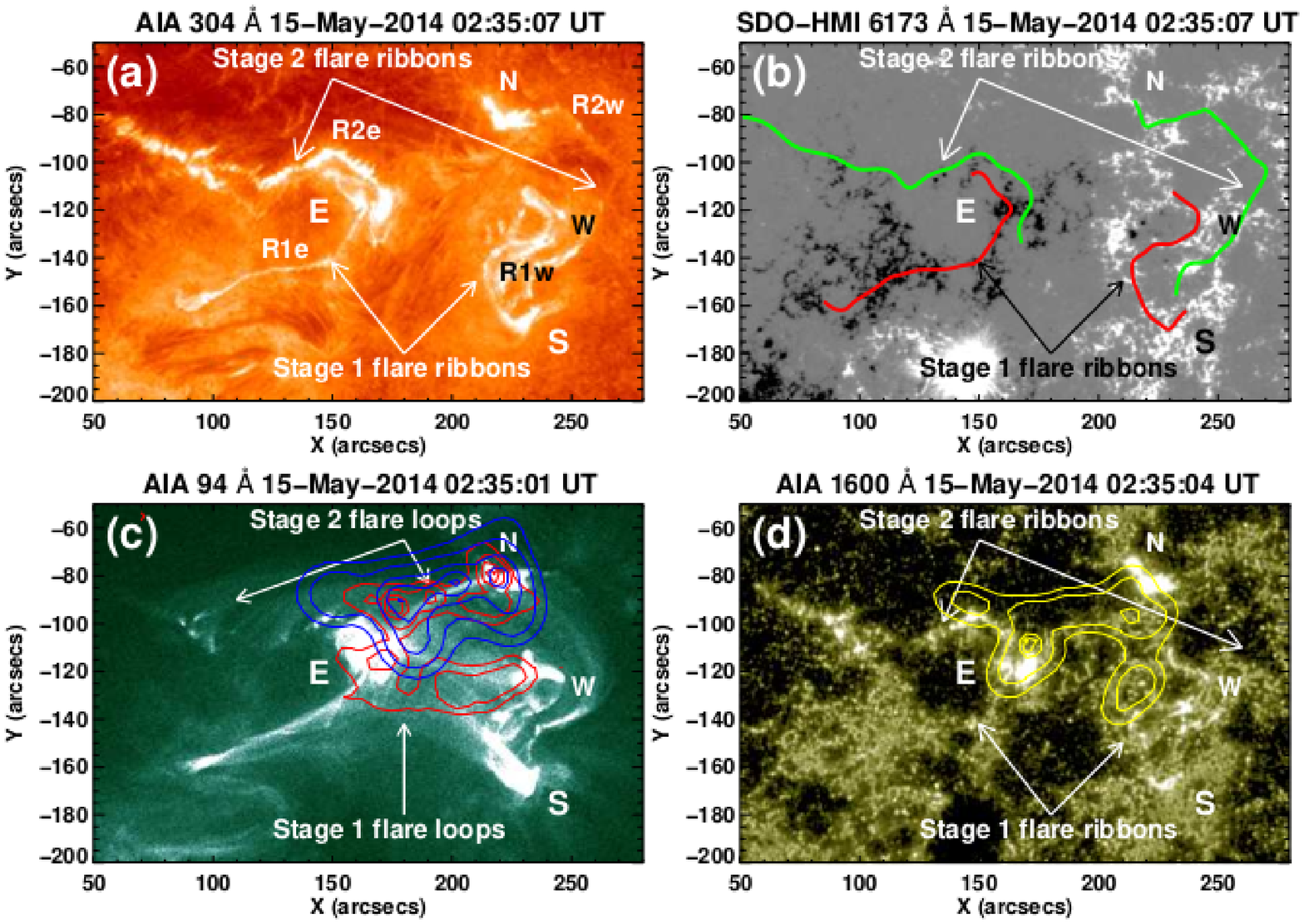}
	}
\vspace*{-6.5cm}
\caption{(a) \textit{SDO}/AIA 304 \AA\ image showing both the first (R1e and R1w) and second (R2e and R2w) set of flare ribbons formed during the stages 1 and 2 of the event, respectively. (b) \textit{SDO}/HMI line--of--sight photospheric magnetogram at 02:35:07 UT on 2014 May 15. The over plotted red and green lines show the first and second set of ribbons, respectively. These ribbons are traced from the AIA 304 \AA\ image that is shown in panel (a). (c) \textit{SDO}/AIA 131 \AA\ image showing the flare loops connecting the flare ribbons. In panel (c) the loops joining the regions E to N are the reconnected loops formed during stage 2 of the flare. (d) The \textit{SDO}/AIA 1600 \AA\ image at 02:35:04 UT, showing the flare ribbon brightenings. The yellow, red and blue contours represent the RHESSI X--ray contours of 3--6, 6--12, and 12--25 keV energy bands. Contours levels are 15\%, 25\%, 50\%, 75\%, and 95\% of the peak intensity.}
\label{fig10}
\end{figure}

In Figure~\ref{fig10}, we show the important multiwavelength features associated with stage 2 of the confined event. We clearly see the different sets of flare ribbons and the flare loops. The newly formed eastern and western ribbons are marked by green colour over the co--temporal magnetogram in Figure~\ref{fig10}(b). For comparison, the ribbons formed in stage~1 are also shown in Figure~\ref{fig10}(b) by red lines. The eastern and western ribbons are located at the negative and positive polarities, respectively. The newly formed flare loop system readily visible in the hot 94 \AA\ channel (Figure~\ref{fig10}(c)), is anchored at ribbons R2e and R2w.

%--------------------------------------------

\subsubsection{Loops in stage 3}
\label{sec3.3.4}

\begin{figure}[!ht]
\vspace*{-8cm}
\centerline{
	\hspace*{0.0\textwidth}
	\includegraphics[width=3.5\textwidth,clip=]{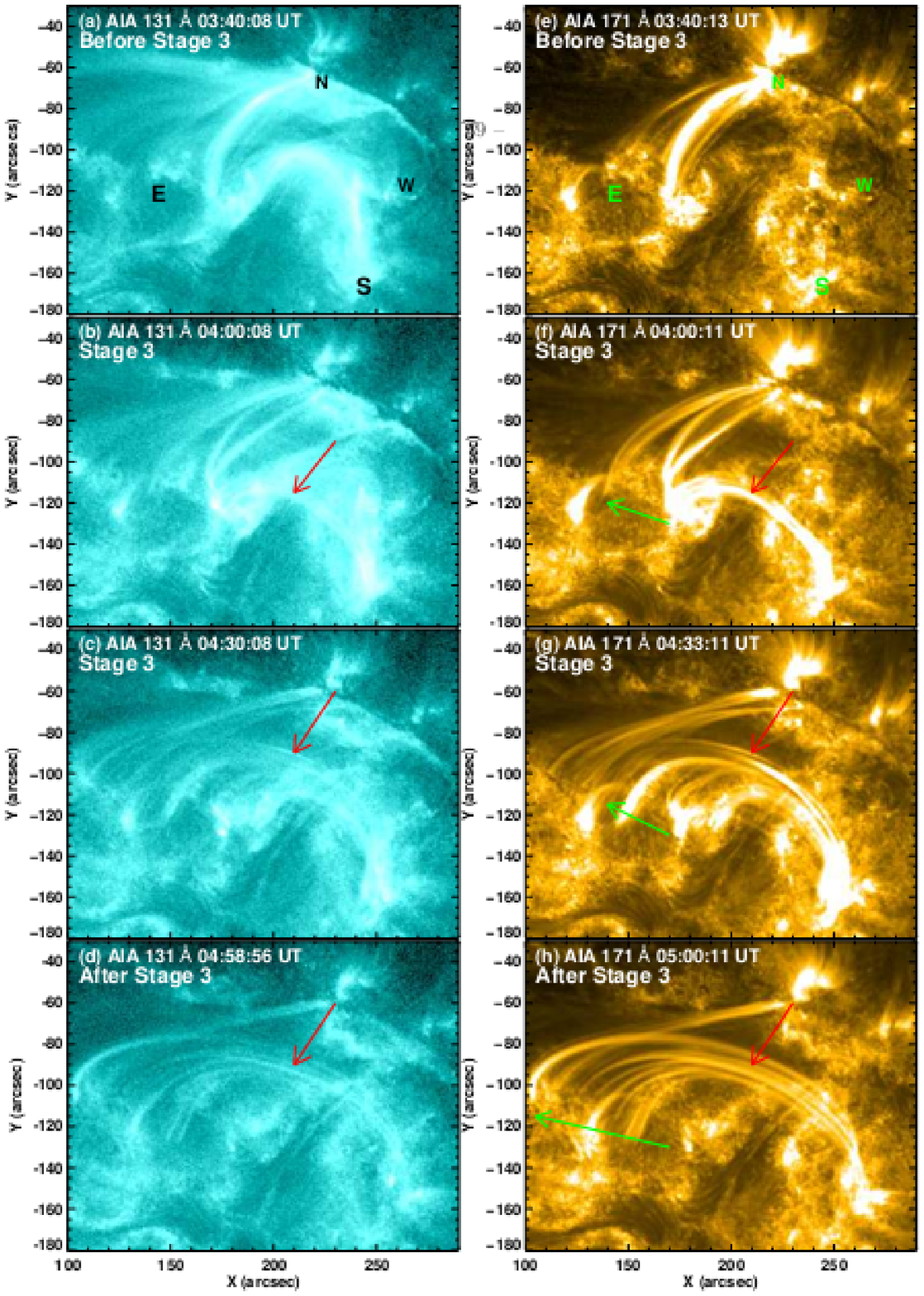}
	}
\vspace*{-7cm}
\caption{The sequence of selected \textit{SDO}/AIA 131 \AA\ (left column) and 171 \AA\ (right column) images, showing the overall flare loops dynamics during stage 3 of the flare event. The right column is available as an electronic animation.}
\label{fig11}
\end{figure}

The evolution of the third stage of magnetic reconnection of the flare and formation of subsequent loop system can be readily observed from the AIA 171 \AA\ and 131 \AA\ images (and corresponding AIA 171 \AA\ movie). The loop system formed following this stage is shown by red arrows in different panels of Figure~\ref{fig11}. We find that the newly developed loops system connects the E and S regions, and  undergoes a rapid evolution in terms of altitude as well as lateral expansion. In particular, the lateral expansion of the loop system toward the east is quite evident (shown by yellow arrows in panels f, g, h).  This is similar to the formation of nearly identical loops during the stage 1 which connect flare ribbons R1e and R1w (Figure~\ref{fig9} and 171 \AA\ movie).  Due to observing limitations, it is hard to ascertain whether parts of the loops system observed during stage 3 developed during stage 1 or not. This uncertainty in distinguishing the formation stages of the two apparently different loop systems at nearly same locations is due to the fact that the AIA 131 \AA\ channel is sensitive to two widely spaced temperature regions, 10 MK and $\rm <1 MK$ \citep[see][]{Boerner12}, the later being similar to the peak temperature of the 171 \AA\ filter. However, here we emphasize that there is a time gap of over 1~hr between stage 1 and stage 3. Furthermore, in AIA 171 \AA\ movie, one can clearly find a complete restructuring of the region in stage 2 during which loops of stage 1 completely disappear. It is noteworthy that although the loops in stage 3 initially share somewhat common footpoints to that of the loops of stage~1, they show distinct lateral expansion toward the east while the western footpoint remains intact. During the lateral expansion of the loop system, the loops show enhanced bright emission. Therefore, we interpret these observations as clear evidence of a new phase of energy release associated with the third stage of magnetic reconnection. 

%--------------------------------------------

\subsubsection{RHESSI observations}
\label{sec3.4.5}

RHESSI observations are explored to investigate the spatial changes in the magnetic reconnection site as well as the temporal characteristics of the energy release during various phases of this atypical confined flare. The RHESSI covered the rise and maximum phases of the flare (i.e., stages 1 and 2) while it missed part of the decay phase (i.e., stage 3). It is worthwhile to note that RHESSI observed the rise phase of the event with the A0 attenuator state, i.e., the observations are recorded with the highest sensitivity. For the RHESSI image reconstruction, we have used the computationally expensive PIXON algorithm \citep{Hurford02} and selected detectors 3F--9F with an integration time of 20 s.

\begin{figure}[!ht]
\vspace*{-7cm}
\centerline{
	\hspace*{0.0\textwidth}
	\includegraphics[width=2.7\textwidth,clip=]{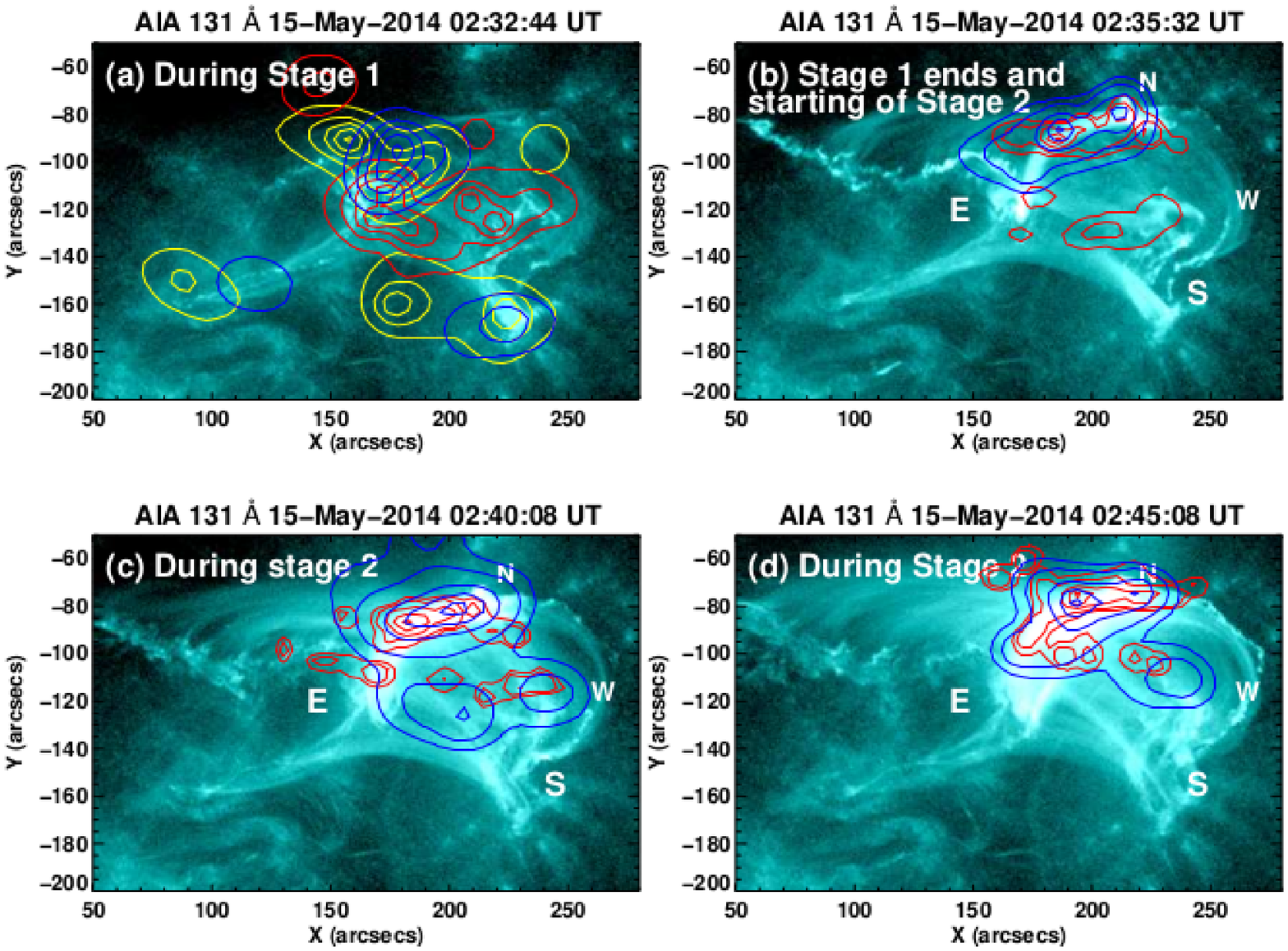}
	}
\vspace*{-7cm}
\caption{((a)--(d)) RHESSI X--ray sources overplotted on the \textit{SDO}/AIA 131 \AA\ images. The yellow, red and blue contours represent the RHESSI X--ray contours of 3--6, 6--12, and 12--25 keV energy bands. Contours levels are 15\%, 25\%, 50\%, 75\%, and 95\% of the peak intensity.}
\label{fig12}
\end{figure}

We find that during stage 1, there is an X--ray source at 6--12 keV energies that connect the region E with regions S and W (Figure~\ref{fig9}(c)). Notably, this source presents two clear centroids that lie over the closest parts of the conjugate flare ribbons besides emission from hot loops formed in--between these ribbons (see also Figure~\ref{fig12}(a)). The bright, distinct centroids provide clear evidence for the particle acceleration at the foot points of the loop arcade, in response to the coronal magnetic reconnection associated with stage 1 \citep[see, e.g.,][]{Joshi11,Guo12,Joshi17}. During the second phase ($\approx$02:35 UT), we find a newly formed group of X--ray sources at various energies (i.e., 3--6, 6--12, and 12--25 keV) that are concentrated in the regions between E and N (Figure~\ref{fig10}(c)--(d)). Furthermore, at this stage also, we find distinct centroids at energies $\rm >6~keV$ that are cospatial with the newly formed flare ribbons in the region, signifying the onset of magnetic reconnection during stage 2. In Figure~\ref{fig12}, we compare the X--ray emissions at various energies during different epochs of stages 1 and 2. Changes in the morphological structure of X--ray sources between stage 1 and stage 2 clearly reveal the onset of magnetic reconnection at the northern region associated with the E--N loop system during stage 2 (c.f. Figure~\ref{fig12}(a)--(b)). We find that at higher energies, the sources were concentrated over and in--between the ribbons (i.e., the main flaring sites) while at lower energies $\rm (<6~keV)$ the emission originated from the main flaring sites as well as away from it. The low energy emission away from the flaring site implies the presence of heated plasma in the active region and could be observed due to the A0 attenuator state of the RHESSI.

%------------------------------------------------------------------------------------------
\section{Magnetic field modeling}
\label{sec4}

%------------------------------------------------------------------------------------------
\subsection{NLFFF extrapolation}
\label{sec4.1}

An NLFFF field extrapolation was performed using an MHD relaxation method described in \cite{Zhu13,Zhu16}, to compute the magnetohydrostatic state of the solar atmosphere. We adopt the computational domain of $709 \times 612\times 258~$Mm$^{3}$ with a resolution of about $\rm 720~km~pixel^{-1}$ of binned data of \textit{SDO}/HMI vector magnetograms \citep{Hoeksema14}. The vector field is obtained through the Very Fast Inversion of Stokes Vector \citep[VFISV;][]{Borrero11} which is a Milne--Eddington based algorithm. A minimum energy method \citep{Metcalf94,Leka09} is used to resolve the $\rm 180^{\circ}$ ambiguity in the transverse field.

Magnetic field modelling is required to analyze the magnetic topology of the AR associated with the flare event. For eruptive flares, it is necessary to follow the evolution of the magnetic field configuration because a CME leads to complete restructuration of coronal magnetic fields \citep{Savcheva2012,Savcheva2016,Janvier16}. On the other hand, the magnetic field configuration does not change drastically during confined flares. Therefore, single magnetogram can be used for fitting the observed flare loops involved before and after the flare \citep{Mandrini96,Dalmasse15,Zuccarello2017,Green2017}.

\begin{figure}[!ht]
\vspace*{-8cm}
\centerline{
	\hspace*{0.0\textwidth}
	\includegraphics[width=3.6\textwidth,clip=]{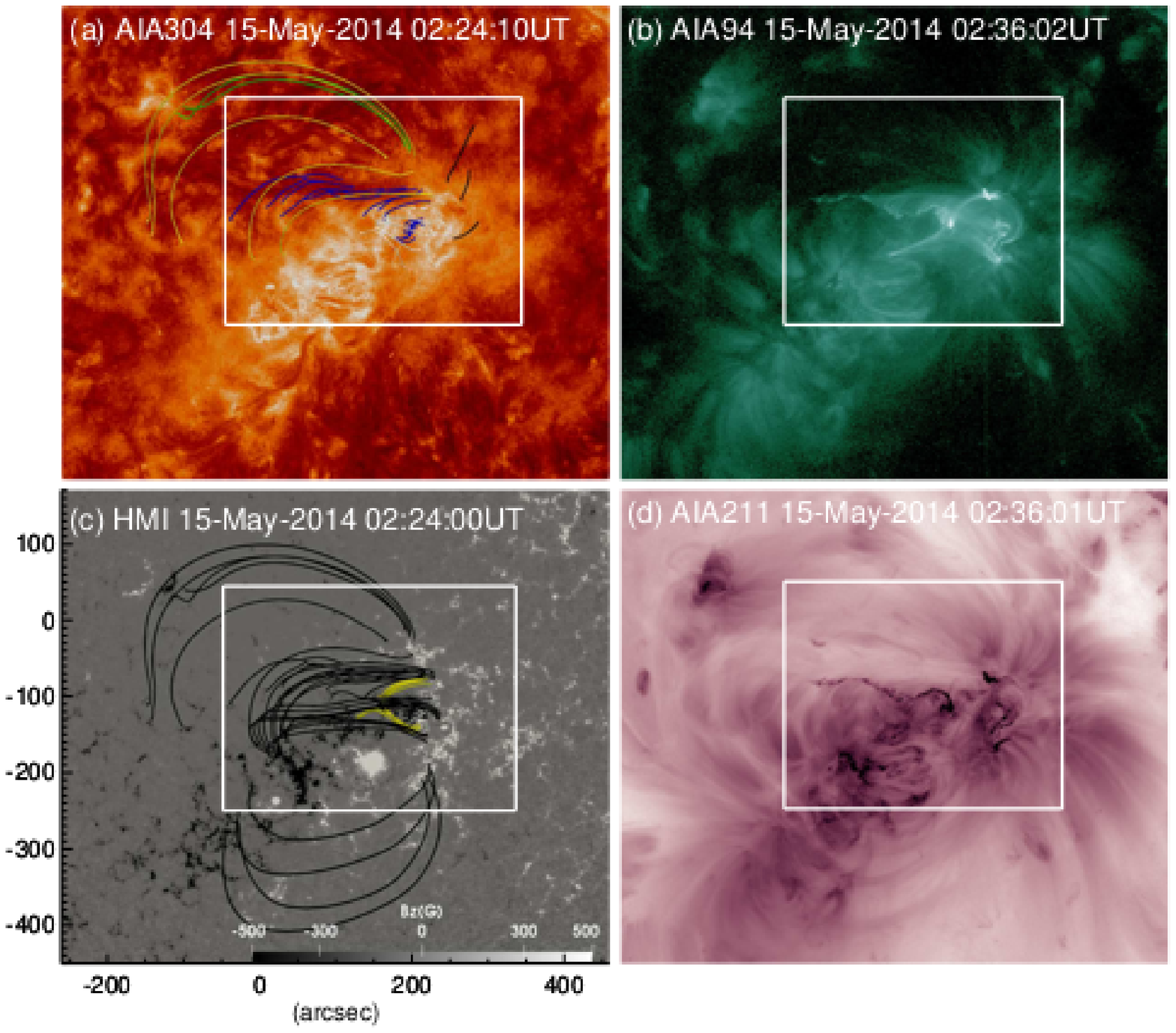}
	}
\vspace*{-10cm}
\caption{(a) \textit{SDO}/AIA 304 \AA\ image overplotted with the selected field lines from the magnetic field extrapolation introduced in Section~\ref{sec4.1}. Blue lines are the small arcades of the two filaments. Yellow and white lines are larger arcades of the filament F1 and F2, respectively. (c) \textit{SDO}/HMI line--of--sight magnetogram with the overplotted modeled field lines. Yellow lines correspond to the bright loops observed during the flare reconnection at AIA 94 \AA\ wavelength image in panel (b). \textit{SDO}/AIA 94 \AA\ (b) and 211 \AA\ (d) images at $\approx$02:36 UT. The reverse colour table is used in AIA 211 \AA\ image (panel (d)). The white frame outlines the field of view of Figure~\ref{fig15}.}
\label{fig13}
\end{figure}

In the present case, we confirmed that, before, during, and after the flare, the large--scale topology did not change, hence, we concentrate our study with the magnetogram obtained before the flare at $\approx$02:24 UT on May 15. Figure~\ref{fig13} shows the overall magnetic field configuration of the active region and filament environment. These figures show that with the NLFFF extrapolation, we are able to fit correctly the loops that we have observed, e.g., the large ones which are nearly in a potential state and the more sheared small flare loops. In the panels (a) and (c) of Figure~\ref{fig13}, we draw the extrapolated magnetic field lines over AIA 304 \AA\ image and HMI magnetogram, respectively. We find that the modelled magnetic field lines drawn by the black colour in Figure~\ref{fig13}(c) nicely match with bright large loops observed in the AIA 211 \AA\ images (Figure~\ref{fig13}(d)). Furthermore, the extrapolated short field lines that cross the PIL, denoted by yellow lines in Figure~\ref{fig13}(c), show a good match with the hot short loops observed during the flare reconnection in the AIA 94 \AA\ images (Figure~\ref{fig13}(b)). In Figure~\ref{fig13}(a), different coloured lines represent various sets of field lines that we have identified for our schematic representation (Figure~\ref{fig16}) and subsequent physical interpretation. It is noteworthy that blue lines in our representation (Figure~\ref{fig13}(a)) essentially denotes low--lying field lines that lie over the two PILs associated with magnetic field systems of filaments F1 and F2 (c.f. Figures~\ref{fig13}(a) and~\ref{fig3}).

\begin{figure}[!ht]
\vspace*{-10cm}
\centerline{
	\hspace*{-0.03\textwidth}
	\includegraphics[width=2.5\textwidth,clip=]{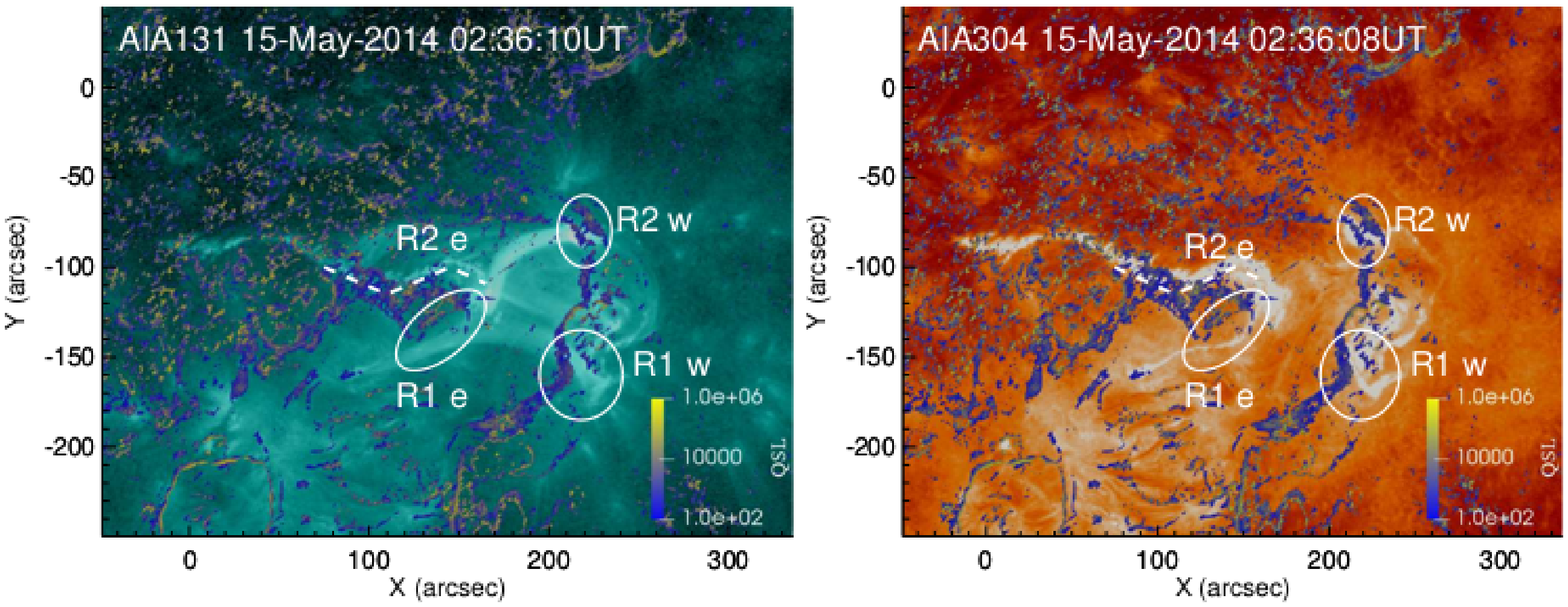}
	}
\vspace*{-8.5cm}
\caption{\textit{SDO}/AIA 131 \AA\ (left panel) and 304 \AA\ (right panel) images with overplotted Q map $(10^{2}<Q<10^{6})$. White ovals show the areas where the QSLs are matching with the flare ribbons.}
\label{fig14}
\end{figure}

The first aim of doing the  NLFFF extrapolation was to find anti--parallel loops which can reconnect as per the scenario of the standard 2D model of solar flares. No such sets of field lines could be identified. Next, we wanted to detect the existence of a null point over the region as it is frequently observed in case of confined flares \citep{Masson09,Guo12,Mandrini2014,Joshi15,Joshi17,Zuccarello2017}. In our case, no high altitude coronal null point has been detected. Finally, the last solution to understand this flare was to make a topological analysis by computing the quasi--separatix layers in order to find the field lines involved in the reconnection process. We will come back to a precise description of the loops that are presumably involved from pre- to post--flare phases of this complex event after analyzing the quasi--separatrix topology of the active region. The QSL position will guide our selection of the field lines before and after the reconnection.

%------------------------------------
\subsection{Quasi-separatrix layers}
\label{sec4.2}

QSLs are defined as robust thin volumes in the corona indicating where the magnetic field gradient is the strongest. They generalize the concept of separatrix and separators in 3D \citep{Demoulin96,Demoulin97}. The footprints of QSLs are the locations where the connectivity of field lines changes drastically. This means that field lines anchored at a polarity in a neighbouring area will see their opposite footpoints anchored at drastically different locations. QSLs are preferential locations for strong electric currents to arise \citep{Aulanier05,Janvier13}. The squashing degree Q is a parameter which indicates the gradient of connectivity change in the magnetic field volume  under consideration \citep{Titov02}. As high squashing factor Q indicates the degree of magnetic field distortion, and therefore an increased probability of finding high electric current densities, the regions of high--Q are associated with locations where magnetic reconnection is the most likely to take place. As QSLs are properties of a large--scale magnetic field volume, extrapolations provide a mean to calculate the locations of these QSLs, and hence of potential reconnection sites. However in many case studies, a one--to--one perfect match between the QSL footprints and ribbons is difficult to achieve \citep{Dalmasse15,Zuccarello2017,Green2017}. This is because QSLs are estimated from magnetic field extrapolations that have their own limitations, while flare ribbons are related with particle beams and chromospheric dynamics. Nevertheless, it is possible locally to match them in some cases e.g., for flares with circular ribbons \citep{Masson09,Janvier16}. \cite{Dalmasse15} demonstrated the relative robustness of the QSL location even if they are computed with an LFFF extrapolation. In the present work, we used a NLFFF extrapolation, so that the extrapolated magnetic field connectivity does depend, to some extent, on the existence of the two filaments F1 and F2 in the active region.

\begin{figure}[!ht]
\vspace*{-9cm}
\centerline{
	\hspace*{-0.05\textwidth}
	\includegraphics[width=3\textwidth,clip=]{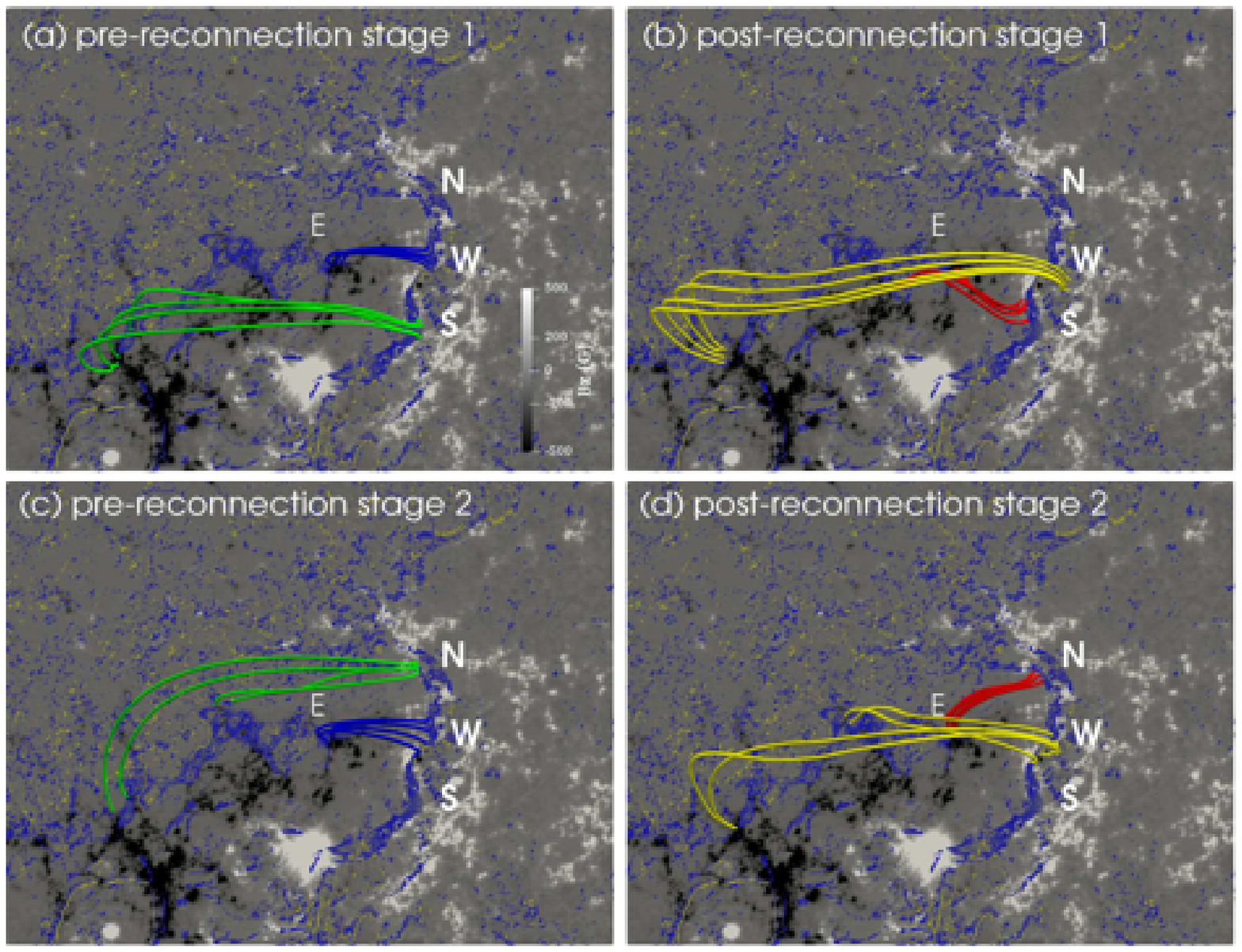}
	}
\vspace*{-9cm}
\caption{\textit{SDO}/HMI line--of--sight photospheric magnetogram overplotted with the modeled field lines and Q map $(10^{2}<Q<10^{6})$, shows the selected field lines and magnetic topology change during the first ((a)--(b)) and second ((c)--(d)) stages of the flare event. Green and blue lines are the pre--reconnected lines, while the red and yellow lines are the post--reconnected lines, respectively.}
\label{fig15}
\end{figure}

Figure~\ref{fig14} displays the \textit{SDO}/AIA 131 \AA\ and 304 \AA\ images at $\approx$ 02:36 UT for a detailed visualization of flare ribbons. QSL values computed from the extrapolation of the magnetogram at $\approx$ 02:24 UT are overplotted. The circles and ellipses represent the areas where the QSLs best match with the ribbons. Since the region is mostly facular, and as such the vertical photospheric current density $\rm Jz$ is mostly noisy, it is difficult for the NLFFF extrapolation to recover exactly the coronal currents. Therefore, we do not expect to have a perfect co--alignment but this qualitative match suggests that the extrapolation reproduces the area well.

As such, the calculation of the squashing factor Q and therefore the QSLs show some mismatch when compared with the locations and the morphology of the flare ribbons. For example, the ribbon R2w is displaced compared to the north--south QSL footprint on the right in the map, and the arc--shaped ribbon between R2w and R1w is not matched with an equivalent QSL footprint. The mismatch of the QSLs and the ribbons come from the weak values of the magnetic field in the environment of the filaments which lead to limitations for the NLFFF extrapolations. 

For instance, the ribbon R1w in stage 1 further extends towards the north during stage~2 forming R2w but the QSLs do not move further west, closer to the new ribbon R2w, even when we used a later magnetogram. The reason is that the two ribbons are only separated by a network cell with low magnetic polarity across which connectivity, hence QSL location, can be very sensitive to the coronal electric currents. Similarly, the eastern part of the elongated flare ribbon R2e cannot be retrieved because the western end of filament F1 lies in a very weak field region. The NLFFF extrapolation cannot retrieve such regions that form the boundaries between two systems of loops in a weak field. That being said, the overall shape of the QSLs, as well as certain features of the QSLs still provide a very good correspondence with the ribbon observations. For example, at $\approx$2:36 UT, the westward QSL has the same elbow angle as the R1w bright ribbon and the R2e has a zigzag shape (shown by white dashed line) as the QSL in the east--west direction. We find that the correspondence between the QSL and flare ribbons is acceptable in the region where the prominent flare loop system existed (i.e., the region separated in between the locations E, N, W, S; see, Figure~\ref{fig12})). Such locations are indicated by the white ovals in Figure~\ref{fig14}. A good correlation between the QSLs and the flare ribbons ensures our analysis of magnetic connectivity presented in the Sections~\ref{sec4.3} to be reliable.

\begin{figure}[!ht]
\vspace*{-3cm}
\centerline{
	\hspace*{0.0\textwidth}
	\includegraphics[width=1.6\textwidth,clip=]{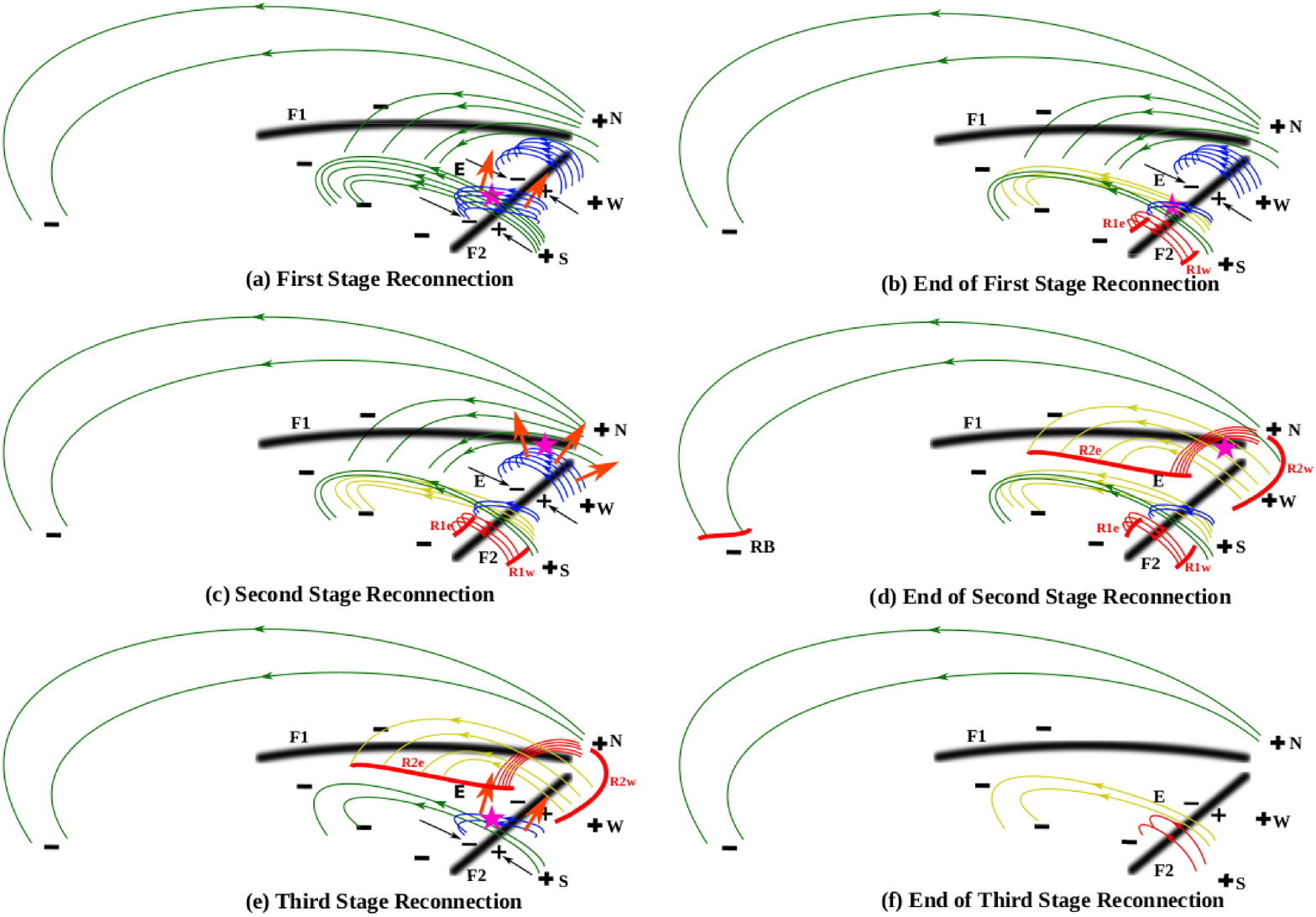}
	}
\vspace*{-2cm}
\caption{Schematic representation of the different stages of the flare event, showing the magnetic topology changes. Green and blue lines are the pre--reconnected lines, while the red and yellow lines are the post--reconnected lines, respectively. Pink stars show the regions of magnetic reconnection.  Changes in the magnetic field lines during stage 3 are identical to that of stage 1.}
\label{fig16}
\end{figure}

%------------------------------------

\subsection{Magnetic field line reconstruction during reconnection}
\label{sec4.3}

We have identified the flare ribbons in AIA 304 \AA\ and the bright flare loops in the hot AIA 94 \AA\ and 131 \AA\ filters (see Section~\ref{sec3}). In Section~\ref{sec4.1}, we found many magnetic field lines which globally fit the observed active region loops. In order to understand which magnetic field line reconnects with another one, we employ a method based on the relationship between QSLs and flaring loops which has been applied in many studies \citep{Mandrini96,Demoulin97,Schmieder97}. 

In Figure~\ref{fig15}, we present the results of the analysis where the set of preflare field lines are in blue and green while the set of post--flare field lines are in red and yellow. To obtain these sets of field lines, we start with the observed bright loops for each stage of the flare. We plot a series of typical field lines rooted at the edges of the QSLs where these bright flare loops are observed. This defines two footpoints per field line.  For example, the bright observed loops ES in stage 1 (Figure~\ref{fig8}(a)) correspond to the red lines of the panel (b) in Figure~\ref{fig15}. Two footpoints of these field lines are defined at E and S locations which are located on one side of the QSL. Then we start from point S but on the external side of the QSL and integrate one of the two field lines that were involved in forming these red loops by reconnection. It goes towards the east, they are represented by the green field lines of Figure~\ref{fig15}(a). The long field line connecting the east side to W corresponds reasonably to the observed long loops from location E. On the other side of the QSL, we find a short field line joining the QSL on the west and that arrive to the point W.  It is represented by the blue lines in Figure~\ref{fig15}(a). The blue and green field lines give us footpoints of two flare loops involved in the reconnection. We, therefore, recover the four points. For stage 2, the same procedure is applied. We start from the red magnetic field line in Figure~\ref{fig15}(d) and finally obtain the two sets of pre- and post--flare loops. For stage 3, the reconnected loops are observed between the locations E and S (see Figure~\ref{fig11}(g) and~\ref{fig11}(h)) which are expected to be similar with the loops identified in stage 1 (see Figure~\ref{fig15}(b)). This is because the QSLs should have moved slightly, while this motion is not shown by our unique extrapolation.

\begin{figure}[!ht]
\vspace*{-3cm}
\centerline{
	\hspace*{0.0\textwidth}
	\includegraphics[width=1.7\textwidth,clip=]{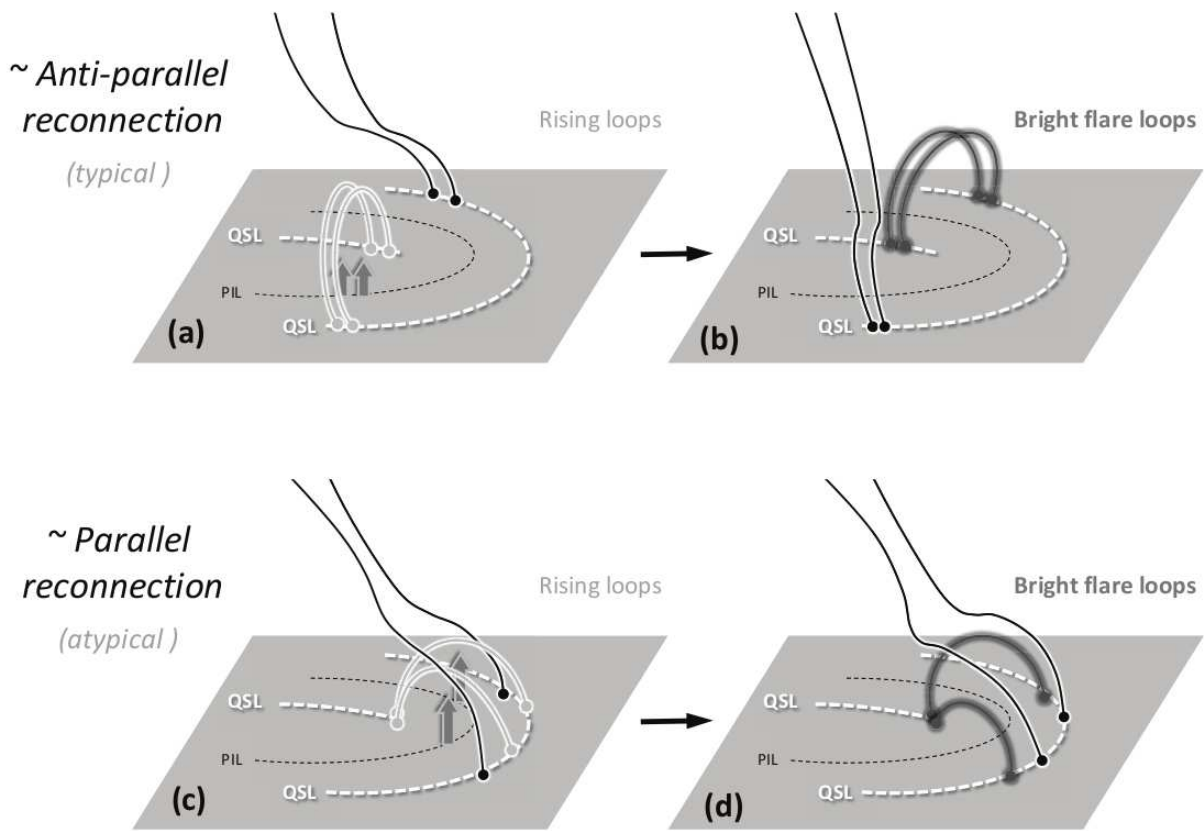}
	}
\vspace*{-2cm}
\caption{Cartoon summarizing the magnetic configuration of compact flares with typical quasi--anti--parallel ((a)--(b)) and atypical quasi--parallel ((c)--(d)) reconnection. The flare presently studied is of the later type.}
\label{fig17}
\end{figure}

In order to understand how the flare was initiated and how it progressed, we propose a cartoon (Figure \ref{fig16}) summarizing the evolution of the magnetic field configuration obtained from the NLFFF extrapolation. Note that in this figure, we have respected the color scheme used for the different set of flare loops in Figure~\ref{fig15} for an easy comparison. There exist two nearby filaments F1 and F2 with common footpoints anchored at the same polarity region, which are overlaid by a low--lying loop system (blue lines in Figure~\ref{fig13}). The flux cancellation underneath the filament  F2 (c.f. Figure~\ref{fig5}) presumably led to its formation, and to the  gradual expansion of its overlying loops (blue loops; Figure~\ref{fig15}(a) and~\ref{fig16}(a)) as shown by MHD simulations \citep{vanBallegooijen89,Aulanier10}. As the expansion progressed, a set of blue loops began to reconnect with the green overlying lines (see Figures~\ref{fig15}(a) and~\ref{fig16}(a)): this is the first stage of the flare. The reconnection region associated with stage 1 is shown by the pink star in Figure~\ref{fig16}(a). As a result, the blue loops and the green overlying lines changed their connectivities and formed the flare loops (red color) and the yellow overlying lines (Figure~\ref{fig15}(b) and~\ref{fig16}(b)). The flare loops and ribbons (marked as R1e and R1w in Figure~\ref{fig9}) are shown in Figures~\ref{fig15}(b) and~\ref{fig16}(b). The expansion of overlying arcades associated with the filament F2 continued in the later phase (see blue lines  in Figure~\ref{fig16}(c)).

In the second stage, the northward expanding arcades of  the filament F2 reconnected with the long loops over the filament F1 (c.f. Figures~\ref{fig15}(c) and \ref{fig16}(c)). The field lines changed their connectivity, and a new set of loops formed on the northern side of the flaring region (red flare loops and the long yellow loops; Figures~\ref{fig15}(d) and \ref{fig16}(d)). The newly formed post-reconnected yellow field lines shown in Figure~\ref{fig16}(d) may also include slipping loops. These loops can be identified in the observations (Figures~\ref{fig6}(b)--(c),~\ref{fig7}(b)--(c) and~\ref{fig10}). Following the reconnection in stage 2, described above, the ribbons (namely R2e and R2w) were formed subsequently (c.f. Figures~\ref{fig15}(d), \ref{fig16}(d), and~\ref{fig10}).

During the third phase, the rearrangement in magnetic connectivity occurred again as we observed long loops joining region E to the western regions W and S (see Figure~\ref{fig11}). Considering that all the loops are quasi-parallel, the energy release in the reconnection cannot be strong and could naturally explain why the flare was weak.

%------------------------------------------------------------------------------------------

\section{Results and discussions}
\label{sec5}

In this work, we present the case study of a small and non--eruptive C class flare that showed significant departure from a ``classical" confined flare in terms of long duration of energy release and development of multiple flare ribbons. The event occurred in composite ARs 12058 and 12060 and involved loops systems across the edges of two adjacent filaments F1 and F2 that remained undisturbed during the flare energy release. The comprehensive analysis of the flare in (E)UV wavelengths and their comparison with modelled coronal magnetic field configuration by a NLFFF technique reveal successive stages of reconnection involving magnetic field lines overlying the two neighbouring filaments. The spatial and temporal evolution of the RHESSI X--ray emission during the rise and maximum phases of the flare provide clear evidence that the magnetic reconnection sequentially proceeded at two adjacent locations within the core region of the flaring environment.

Flux cancellation and photospheric motions have already been observed to be the main sources for compact flares \citep{Hanaoka97,Schmieder97,Nishio97,Chandra06,Dalmasse15}. Here, we observed the formation of filament F2 from 2014 May 13 to May 15 during the continuous flux cancellation in the region underneath the filament (see Figures~\ref{fig4} and~\ref{fig5} and associated animation). During the formation of flux rope  F2, we assume that the loops in the active region should evolve as shown in the model of \citet{Aulanier10}. The expanding loops overlying filament F2 reconnect with the surrounding field as well as with the overlying loops of filament F1 successively in three different phases. Notably, the spatial evolution of the X--ray footpoint sources from E--W to E--N directions, with the relatively small spatial shift in the eastern footpoint gives credence to the interpretation above, in which magnetic reconnection successively involves adjoining loop systems.

Confined/compact flares ensue when the reconnection occurs between two groups of loops \citep[e.g.,][]{Hanaoka97} that must involve QSLs \citep{Mandrini96,Schmieder97,Demoulin97} or at a coronal null point with a single spine that emerges away from the fan surface anchored in a remote region \citep[e.g.,][]{Masson09,Hernandez-Perez17}. Confined flares are also observed in the active region where the overlying magnetic field is too strong to allow the filament (flux rope) to erupt \citep[e.g.,][]{Amari99,Torok05,Guo10,Kushwaha14,Joshi14a,Sun15,Kushwaha14,Kushwaha15}. 

In the present work, we want to generalize the magnetic configuration conditions in which long duration confined flares can occur without eruptions. Figure~\ref{fig17} summarizes different typical conditions for getting the interaction between adjacent loops. Recently, \cite{Dalmasse15}, studied a similar type of atypical compact flare and found out the interaction between different loop systems from a deep analysis of QSLs.  We also computed the QSLs to find a match with the observed ribbons which are the footpoints of the arcade loops. The scenario is different from the standard models of solar flares, where the reconnection occurs underneath the erupting filaments among the legs of erupting arcades. Differently, here we observe the formation of flare arcades over the filaments systems (see Figure~\ref{fig6}), without any disturbance and eruption of either filament. In the flare scenario, we also believe that the moving bright kernels seen during the flare are the signatures of slipping, and maybe slip--running reconnection, as this process is intrinsic to 3D reconnection at the core of the flaring mechanism \citep[see][]{Aulanier06}.

We summarize our scenario for such compact flare configurations in Figure~\ref{fig17} after comparing our case with more usual set-ups used for compact flares from extrapolation \citep{Mandrini96,Schmieder97,Demoulin97} and MHD simulations \citep{Moreno08,Pariat09,Torok09}. Commonly, we get the same ``usual" 3D configuration in various confined flare events with a strongly curved PIL and a straight QSL footprint surrounded by a horse--shoe (or arc--shaped) QSL footprint. Depending on the location of the rising loops (that can arise due to flux-emergence, or underlying flux rope build--up like in our case), we can get:\\
- quasi--anti--parallel reconnection leading to one set of flare loops (as commonly observed)\\
- quasi--parallel reconnection leading to two sets of flare loops (like in our case)\\
The latter may lead to a false impression of an eruptive two--ribbon flare. The present study, thus adds up a new well--identified case of so--called atypical flares.

After having in hand different case--studies of confined flares \citep{Liu14,Dalmasse15} and the present study that together form an ensemble of well--identified and various situations for atypical flares, we can conclude on the key mechanism occurring in such flares. These events look like two ribbons flares but, in fact, the forcing lies in the interaction of quasi--parallel loops. We show the importance of the curvature of the PIL, that creates a purely 3D topological effect, which in turn allows reconnection between quasi--parallel field lines, and which exists neither in 2D models nor in models of failed eruptions.

%------------------------------------------------------------------------------------------

\acknowledgments
We are grateful to the referee for his/her valuable comments and suggestions that significantly improve the scientific content and presentation of the paper. We thank \textit{SDO}/AIA, \textit{SDO}/HMI, \textit{GOES} and \textit{RHESSI} teams for providing their data for the present study. This work is supported by the BK21 plus program through the National Research Foundation (NRF) funded by the Ministry of Education of Korea. BS thanks Prof. Magara to invite her to Kyung Hee University on the BK21 program where this project started to be discussed. RC and BJ acknowledge the support from SERB-DST, New Delhi project No. SERB/F/7455/2017-17.

%------------------------------------------------------------------------------------------

%\bibliography{reference.bib}
%\bibliographystyle{apj} % style aa.bst
%\bibliography{reference} % your references Yourfile.bib

%-----------------------------------------------------------------------------------------

\end {document}